\documentclass[twocolumn,10pt]{article}
\usepackage{usenix}

\AtBeginDocument{%
  \providecommand\BibTeX{{%
    \normalfont B\kern-0.5em{\scshape i\kern-0.25em b}\kern-0.8em\TeX}}}

\usepackage{booktabs}

\usepackage{amssymb}
\usepackage{flushend, subcaption}
\usepackage{graphicx}
\usepackage{url}

\usepackage{listings}
\usepackage{fancyhdr,graphicx,amsmath,amssymb}
\usepackage{algpseudocode}
\usepackage{algcompatible}
\usepackage[ruled,vlined]{algorithm2e}
\usepackage{enumitem}
\include{pythonlisting}
\algnewcommand\algorithmicreturn{\textbf{return}}
\algnewcommand\RETURN{\algorithmicreturn}
\algnewcommand\algorithmicprocedure{\textbf{procedure}}
\algnewcommand\PROCEDURE{\item[\algorithmicprocedure]}%
\algnewcommand\algorithmicendprocedure{\textbf{end procedure}}
\algnewcommand\ENDPROCEDURE{\item[\algorithmicendprocedure]}%
\algnewcommand{\algvar}[1]{{\text{\ttfamily\detokenize{#1}}}}
\algnewcommand{\algarg}[1]{{\text{\ttfamily\itshape\detokenize{#1}}}}
\algnewcommand{\algproc}[1]{{\text{\ttfamily\detokenize{#1}}}}
\algnewcommand{\algassign}{\leftarrow}

\newcommand{\PP}{PayloadPark}

\newcommand{\PPTTL}{EXP}

\newcommand{\nonPP}{baseline}

\newcommand{\singleindent}{\hskip0.5em}
\newcommand{\doubleindent}{\hskip1em}
\newcommand{\tripleindent}{\hskip1.5em}
\newcommand{\fourthindent}{\hskip2em}

\algrenewcommand{\algorithmiccomment}[1]{$\triangleright$ #1}
\usepackage[labelfont=bf,textfont=bf]{caption}
\usepackage{multirow}
\usepackage{makecell}

\begin{document}

\title{Parking Packet Payload with P4}

\author{
{\rm Swati Goswami, Nodir Kodirov, Craig Mustard, Ivan Beschastnikh, Margo Seltzer} \\
University of British Columbia\vspace{-10cm}
}

\maketitle
\begin{abstract}

Network Function (NF) deployments suffer from poor link goodput,
because popular NFs such as firewalls process only packet headers
while receiving and transmitting complete packets.
As a result, unnecessary packet payloads needlessly consume
link bandwidth.
We introduce PayloadPark, which improves goodput
by temporarily parking packet payloads in the stateful memory of
dataplane programmable switches.
PayloadPark forwards only packet headers to NF servers, thereby
saving bandwidth between the switch and the NF server.
PayloadPark is
a transparent in-network optimization that complements existing
approaches for optimizing NF performance on end-hosts.

We prototyped \PP{} on a Barefoot Tofino ASIC using the P4
language. Our prototype, when deployed on a top-of-rack switch, can service
up to 8 NF servers using less than 40\% of the on-chip memory resources.
The prototype improves goodput by 10-36\% for Firewall and NAT NFs and by
10-26\% for a $Firewall \rightarrow NAT$ NF chain without harming latency.
The prototype also reduces PCIe bus load by 2-58\% on the NF server thanks to
the reduced data transmission between the switch and the NF server.
With workloads that have datacenter network traffic characteristics,
\PP{} provides a 13\% goodput gain with
the $Firewall \rightarrow NAT \rightarrow LB$  NF chain without latency penalty.
In the same setup, we can further increase the goodput gain to 28\% by using packet recirculation.
\end{abstract}

\section{Introduction}
\label{sec:intro}

Network Functions (NFs) are widely deployed in the enterprise network~\cite{APLOMB}.
NFs, such as firewalls and NATs, typically examine only a small part of each packet.
These NFs show poor goodput, because unexamined packet payloads consume valuable
link bandwidth.
\emph{Goodput} is the amount of useful information delivered over time and is
a measure of how effectively a link is used.
In this case, goodput is the amount of data examined by the NF.
For instance, NATs and firewalls can achieve \emph{throughput} that saturates a 40Gbps
link when processing 10Mpps with 500 byte (4000 bits) packets.
But, NATs and firewalls only examine the 5-tuple in the packet header, approximately
only the first 42 bytes\footnote{Including Ethernet,
IPv4, and UDP header.} (336 bits) of the UDP packet.
In these cases, the resulting \emph{goodput} is only 3.36Gbps.
To increase goodput, we propose \PP{} -- a header-payload decoupling
optimization that temporarily holds packet payloads in switch dataplane memory.
\PP{} forwards only packet headers to NFs, temporarily
parks the payload in the switch ASIC memory, and reassembles the packet when
returned by the NFs.

\begin{figure}[t]
  \includegraphics[width=\linewidth]{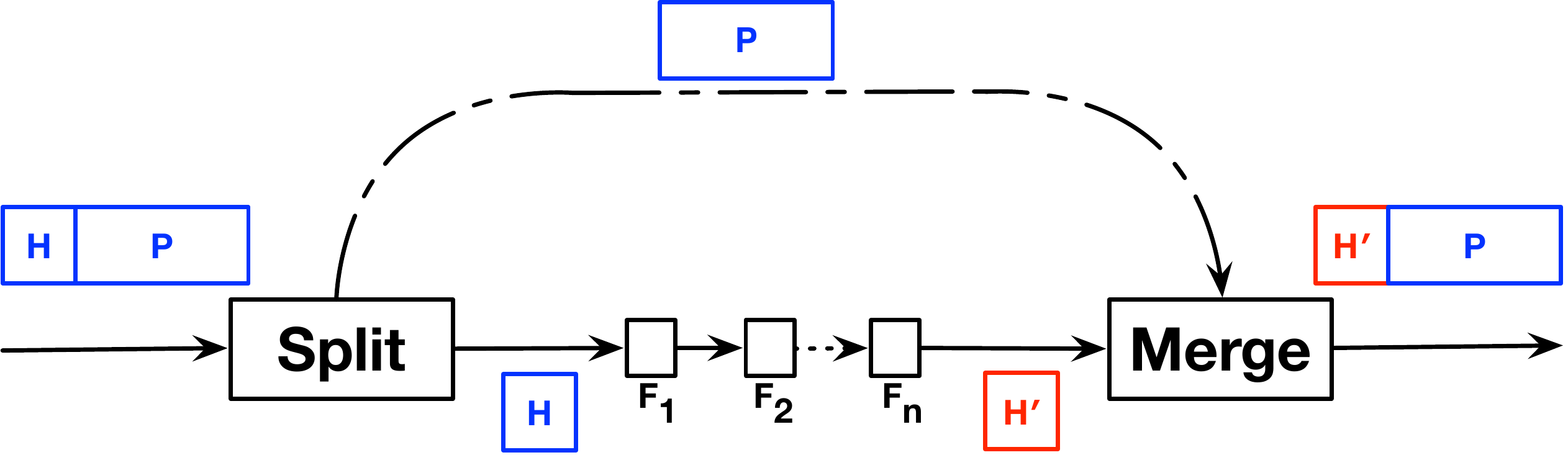}
  \caption{Abstract \PP{} deployment. \textit{Split} decouples the packet into
  header H and payload P. The header H is then processed by a shallow NF chain
  F\textsubscript{1}, F\textsubscript{2} ... F\textsubscript{N}.
  \textit{Merge} reassembles the resulting header H' with payload P.}
  \label{fig:\PP{}_abstract}
\end{figure}

\PP{} presents a unique design point as it uses specialized hardware (RMT
switches) to improve Network Functions (NFs), without requiring NFs to be adapted to
specialized hardware.
NFs implement critical network functionality, such as intrusion detection,
NAT, and performance optimization (caches, WAN
optimizers).
NFs are often connected together in an \emph{NF chain}, such as FW$\rightarrow$NAT~\cite{rfc7665}.
Previously, this functionality was implemented by hardware middleboxes that
are inflexible and have vendor lock-in problems.
NFs address these problems by using software deployed on commodity hardware, but
have worse performance~\cite{etsi-whitepaper, Han2015NFVChallenges},
so research efforts focus on optimizing NFs (\S\ref{sec:related_work}).
Proposals to move NFs to the cloud~\cite{APLOMB} makes performance even
more critical due to additional network hops and encryption overhead.
\PP{} retains the \textit{flexibility} of running software-based NFs on
commodity hardware and
simultaneously provides some of the performance
benefits of specialized hardware using RMT switches.

Fig.~\ref{fig:\PP{}_abstract} shows an abstract deployment of \PP{}. We park the
payload by implementing two operations in the switch dataplane: \textit{Split} and \textit{Merge}.
\textit{Split} decouples the incoming packet's header, H, from its payload, P.
Header H is forwarded to the shallow NF chain $F_{1},F_{2} ...F_{n}$,
while payload P is parked (stored) in the switch dataplane.
\textit{Merge} re-combines the (potentially modified)
header H' from the NF chain with
payload P before forwarding the packet to its destination.
Deploying Split and Merge on the same switch uses
the bandwidth between the switch and the NF servers for useful
information (the headers) rather than unused data (the payload), thereby
improving goodput.

As PayloadPark addresses bandwidth, it is orthogonal to prior work that
optimizes endhosts, such as reducing per-packet CPU cycles (SpeedyBox~\cite{SpeedyBox},  NetBricks~\cite{NetBricks}).
In fact, \PP{} complements such existing end-host optimizations.
For example, a combined setup of \PP{} and NetBricks results
in goodput improvement from \PP{}, and throughput and latency gains from NetBricks.
Also, NF frameworks that use specialized devices such as FPGAs~\cite{clickNP} or programmable
NICs~\cite{metron} are bandwidth bound
and will yield a higher benefit by improving the packet processing rate -- a result of improved link goodput.

While intuitive, PayloadPark has only recently become possible thanks to newly
available Reconfigurable Match-Action Table (RMT) switches~\cite{rmt}.
RMT switches are equipped with programmable ASICs that add programmability into
the switch dataplane.
This creates opportunities for implementing in-network optimizations that were
impossible with fixed-function ASICs.
Prior work proposes offloading application logic to programmable switches~(\S\ref{sec:related_work}), but
\PP{} is a domain-specific optimization, and it is transparent to the application.
Our prototype uses RMT switches to temporarily store 160 bytes from each
packet's payload (we expect to be able to increase this with future switch models).
In the evaluation (\S\ref{sec:eval}), we show that storing this small amount of data per packet is
effective in practice.

\noindent \textbf{Applicability of \PP{}.}
\PP{} is appropriate for header-only NFs, i.e., shallow NFs, such as NATs,
firewalls, and L4 load balancers.
Deep packet inspection, such as intrusion detection, inspects the packet
payload, so \PP{} cannot be applied.
Shallow NF processing is widely performed in many enterprise datacenters.
A survey of 57 enterprise networks shows that the number of middleboxes is
comparable to L2 and L3 routers~\cite{APLOMB}.
A prior survey shows that on average, 44\% of datacenter traffic
requires at least one of L4 load-balancing and NAT operations~\cite{ananta}.
Moreover, with increasingly encrypted traffic~\cite{CostOfHTTPS}, many NFs
are effectively limited to shallow processing, furthering the
applicability of \PP{}.
\PP{} works seamlessly with encrypted payloads, because it does not
inspect payload content.

Prior work, such as Dejavu~\cite{dejavu} and SilkRoad~\cite{Miao2017SilkRoad},
move functionality onto the switch (e.g., offloading entire chains or specific
functionality, such as load balancing).
Such deployments provide high performance but limit flexibility in NF
implementation.
SilkRoad stores active connection state in the switch, so it suffers
performance degradation
when the number of active connections exceed stateful memory resources on the switch~\cite{McCauleyLoadBalancing}.
Our work differs from these approaches
in that \PP{} is a
\textit{transparent in-network optimization} that leaves
NF chains to run on commodity hardware. This retains the \textit{flexibility} in
implementing NF chains and ensures ease of integration with existing NF frameworks such as OpenNetVM~\cite{opennetvm}
and NetBricks~\cite{NetBricks}.

\noindent \textbf{Challenges.}
The design of \PP{} and our prototype implementation address the following
three challenges to implement this intuitive optimization.

\noindent \textbf{1. Transparent operation.}
Shallow NFs, such as NATs, modify packet headers. \PP{} must be able to
re-assemble the packet despite these modifications, \emph{without assistance from
the NF framework.}
\PP{} does not make any assumptions about the NF framework deployed on end-hosts,
and it can be integrated with popular
frameworks, such as OpenNetVM~\cite{opennetvm}, NetBricks~\cite{NetBricks}, and OpenBox~\cite{openbox},
without any functional code changes.

\noindent \textbf{2. Limited memory resources.}
\PP{} needs to temporarily hold payloads, but switch dataplanes have limited storage that is
 shared across multiple ports. For example, 6.4+ Tbps RMT switches
 have 50-100MB of stateful SRAM~\cite{Miao2017SilkRoad}.
Our insight is that the low latency of NFs \emph{significantly limits} the
payload storage required.
Shallow NFs have latency on the order of 10s of $\mu s$~\cite{NetBricks}.
Even if the worst case time-delta between Split and Merge operations
(including NF operation)
is an order of magnitude higher, say 200 $\mu s$,
this requires only 0.8MB of storage to saturate a 40Gbps link,
with a 160-byte payload (and a 42-byte UDP header).
In addition, \PP{} works within memory limits by 1) evicting payloads after a
predefined expiry threshold, and 2) falling back to non-\PP{} mode when storage
is exhausted.

\noindent \textbf{3. Limited packet processing time budget.}
RMT switches process packets at line rate by imposing an upper-limit on
the number of compute and storage operations that can be executed per
packet.
This means that there is a limited time budget to find empty space in the switch
memory.
We address this by exploiting the fact that packets are processed by NF
frameworks in (mostly) FIFO order and
falling back to non-\PP{} mode in the uncommon case that empty space cannot be found.

To summarize, we make the following contributions:
\begin{itemize}[topsep=0pt]
    \item We present the design of \PP{} that uses RMT switches to store payload.
    \item We implement and evaluate a prototype of \PP{} using a Barefoot Networks' Tofino ASIC.
    \item We quantify the efficacy of \PP{} to identify under what conditions it provides performance benefits.
\end{itemize}

\section{Background: RMT Switches}
\label{sec:background}


Commodity switches with fixed-function ASICs use pre-configured header
definitions and packet processing logic.
This makes it slow, if not impossible, to evolve the network to support new protocols.
RMT switches address this shortcoming by providing a programmable
dataplane, which puts network engineers in control of header definitions and
packet processing logic.


At a high level, RMT switches work by passing each packet through a series of
match-action tables (MATs), which perform \textit{actions} (functions) on packet
headers that \textit{match} criteria.
More precisely,
the packet processing pipeline consists of three building blocks:
a Parser, Match-Action Pipeline, and Deparser.
The \textit{Parser} interprets the packet header using a \emph{user-defined} header and
populates the Packet Header Vector (PHV) that makes the parsed fields available to
the match-action pipeline.
The PHV also includes user-defined metadata fields, used for passing information
to subsequent MATs.
The maximum PHV size is switch-specific and limits the header size on which the
match-action pipeline can operate.

The \emph{Match-Action Pipeline} is composed of stages, where each stage has
local ALU, SRAM, and TCAM resources.
The match-action pipeline is programmed by writing MAT definitions in
P4~\cite{Bosshart2014P4}.
MATs are mapped to stages by the P4 compiler, and independent MATs can be
mapped to the same stage.
Each MAT contains pairs of \textit{match} rules and \textit{actions}.
For each packet, a MAT compares header fields (from the PHV) using user-supplied \textit{match} rules,
and executes an \textit{action} based on the comparison result.
Action definitions consume hardware resources called Very Large Instruction
Word (VLIW) actions~\cite{rmt}.
Actions can modify PHV contents, add/remove header fields, and route/drop
packets.
For example, an L2 router matches packets on the destination MAC address
and executes an action to forward the packet to the correct switch port.
Finally, the \emph{Deparser} reassembles the new header from the modified and unmodified
header fields.

MATs access SRAM reserved for stateful operations using a read/write register API,
which views all of
stateful memory as an array of fixed size bit-vector registers.
This SRAM is separate from the packet buffer memory.
Despite the added programmability, switches process packets at line rate by
imposing restrictions on the number of per-packet stateful memory accesses
(among other things).
The packet processing pipeline can recirculate the packet in the pipeline, which
sends the packet back to the parser.
Recirculation 
increases the number of permitted per-packet header
transformations but introduces a bandwidth and latency penalty.



\section{\PP{} Overview}
\label{sec:overview}

We begin by outlining \PP{}'s goals and providing a general overview.
Then, we dive into implementation details in \S\ref{sec:design} and
\S\ref{sec:implementation}.

\subsection{Design Goals}
\label{sec:design_goals}

\PP{} has to meet the following design goals.

\noindent \textbf{Transparency.}
\PP{} must give cloud providers the flexibility to enable/disable \PP{} as
needed.
To do so, \PP{} must be functionally equivalent to non-\PP{} (baseline) deployments and
require no changes to the traffic source and the NF framework.
Since \PP{} is agnostic to the NF framework, providers are free to
choose NF frameworks that optimize different metrics, such as SLO
guarantees (ResQ~\cite{ResQ}) or latency (NFP~\cite{nfp},
SpeedyBox~\cite{SpeedyBox}).

\noindent \textbf{Operation with limited ASIC resources.}
Programmable switch resources are shared across ports, so \PP{} must work
within the limits of the switch and leave enough room for packet processing on other ports.
For example, on the 6.4 Tbps switch with Barefoot Networks Tofino ASIC, 16 ports share the
compute and storage resources of the same match-action pipeline.
Using \PP{} for a subset of 16 ports sharing the same match-action pipeline must leave
sufficient resources to implement packet processing logic for the remainder of the ports.

\noindent \textbf{Performance.} \PP{} improves goodput, but NFs are latency
sensitive, so \PP{} must not incur a latency penalty.


\subsection{\PP{} Header}

%


\PP{} is enabled on a per-port basis.
When a packet arrives on a \PP{}-enabled port, the Split operation adds the \PP{}
header (shown in Fig.~\ref{fig:SMP_Header}) to track \PP{}-specific state.
The Enable (ENB) bit indicates if the packet payload was successfully stored in the switch.
The opcode (OP) bit distinguishes between Merge and
Explicit Drop operation (discussed in \S\ref{sec:eval_packet_drops}).
The Tag is used to find the payload in the switch dataplane memory; it is composed
of three subparts: the table index, the generation number, and the CRC.
The CRC is used to validate the \PP{} header before merging the stored payloads
with packets returning from the NF server.


\subsection{High Level Algorithm}
\label{sec:high_level_algorithm}

\begin{figure}[t]
  \includegraphics[width=\linewidth]{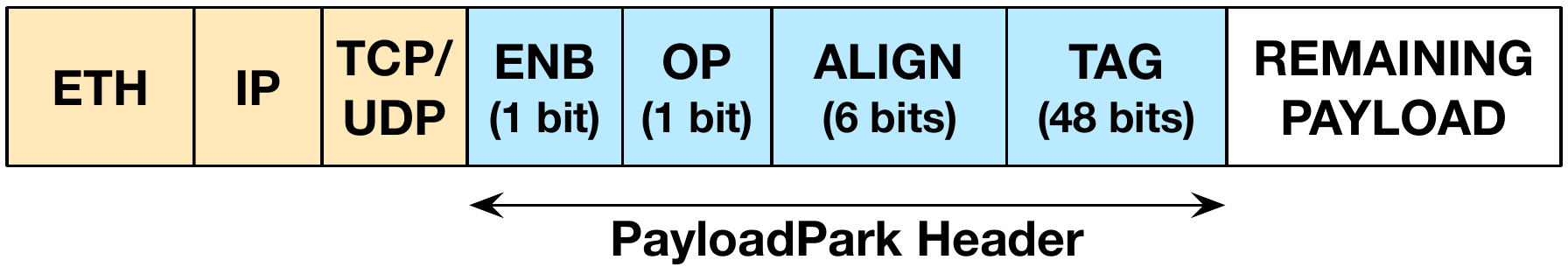}
  \caption{\PP{} header. The Enable (ENB) field indicates whether \PP{}
    operation is enabled for the packet. The Opcode (OP) indicates the
    operation to be performed:  Merge | Explicit Drop.
    ALIGN bits are for byte-alignment of header fields.
    The Tag (TAG) is a unique identifier for
    the packet.
  }
  \label{fig:SMP_Header}
\end{figure}

\begin{figure}[t]
  \includegraphics[width=\linewidth]{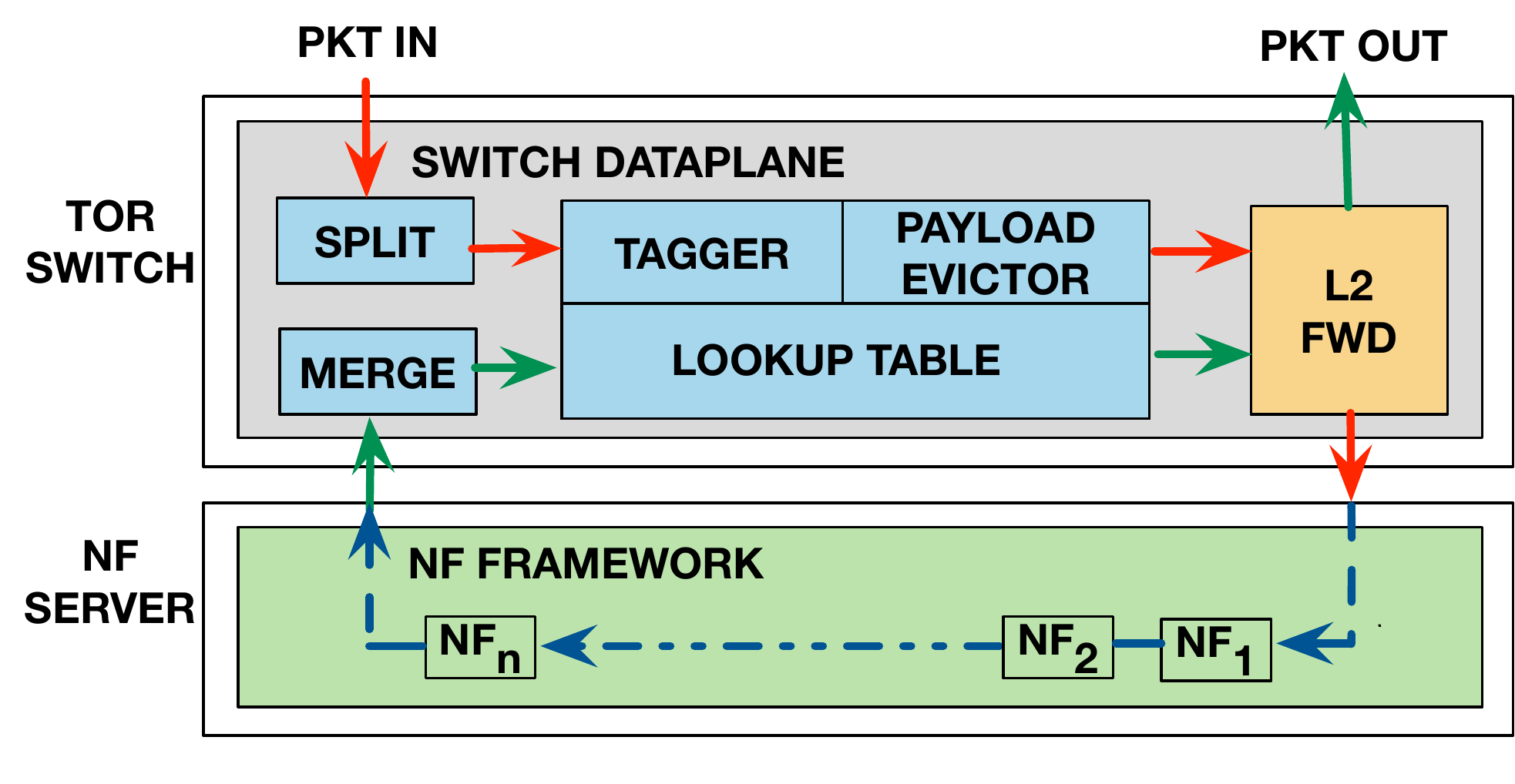}
  \caption{Packet flow in \PP{}.  \textit{Split} decouples the packet header and payload
  and stores the payload in the payload table.
  The \textit{Merge} operation merges the headers from the NF server with the payloads  stored in the lookup table.
  L2 FWD forwards packet using L2 forwarding.}
  \label{fig:SMP_overview}
\end{figure}

\PP{} implements two primary operations: \emph{Split} and \emph{Merge}
as shown in Fig.~\ref{fig:SMP_overview}. This
section describes their high level implementation and their interaction with
auxiliary components. 

\begin{itemize}[topsep=0pt,leftmargin=*]

\item \textbf{Split} decouples the packet header and payload,
by 1) associating a unique tag with the packet,
2) storing the payload in the switch memory,
3) adding the \PP{} header, and 4) setting the Enable bit (ENB bit in Fig.~\ref{fig:SMP_Header}) to one.
If there is insufficient memory in the switch to store
a payload, the Split operation adds the \PP{} header but
sets the Enable bit to zero.

\item \textbf{Merge} recombines the payload and the header.
When the switch receives the potentially modified header
from the NF chain, the Merge operation, 1) uses the tag to locate
the stored payload, 2) appends the payload to the packet,
3) removes the \PP{} header, and 4) frees the
space consumed by the payload.

\end{itemize}

Split and Merge functionality is composed of
three components: packet tagger, lookup table, and payload evictor
as shown in Fig.~\ref{fig:SMP_overview}.

\noindent \textbf{Packet tagger.}  Every \PP{} packet must be assigned a unique identifier
or \textit{tag} that is used to index into the register array.
The packet header cannot be used for indexing in the register array, because NF chains can
modify headers, making it impossible to find the payload.

\noindent \textbf{Lookup table.} \PP{} uses a lookup table abstraction
on top of the raw register API of P4. The lookup table is composed of two tables
-- metadata and payload tables --
each organized as register arrays indexed using a common table index.
The metadata table is conceptually a bitmask indicating which positions
in the payload table are occupied.
The Split and Merge operations allocate and reclaim memory resources in the
lookup table.

\noindent \textit{Allocating memory using the Split operation.}
Split examines the metadata bitmask to find an unoccupied location.
If an unoccupied location is found, Split marks the table entry as occupied, and
stores the payload in the payload table at the same index.


\noindent \textit{Reclaiming memory using the Merge operation.}
Merge examines the Enable bit in the \PP{} header to determine if the packet's
payload is stored in the lookup table.
If so, Merge finds the payload by indexing into the payload table with the tag from the \PP{} header.
Merge adds the payload to the packet, and marks the table entry as empty.

\noindent \textbf{Payload evictor.}
Split packets may get dropped or lost before they return to the switch for merging.
Payloads for packets that never return to the switch consume space and, if left
unchecked, will exhaust the lookup table.
Packet drops may occur when packets are dropped by NFs, such as firewalls, but
since the NF framework is unaware of \PP{}, it will not notify the switch of
such packet drops.
Packet loss can be caused by lossy links and other components.
Dropped or lost packets will never return to the switch, so \PP{} must be able to evict parked payloads to reclaim space on the switch.


We implement payload eviction by augmenting the metadata table with an
expiry threshold.
During the Split operation, this threshold is initialized to a predefined
constant value.
The Split operation decrements the threshold each time it indexes into an occupied location
and evicts the payload when the threshold reaches zero.
We experiment with different threshold values in \S\ref{sec:eval_packet_drops}.


Payload eviction is necessary to reclaim space, but it can cause packet loss
if payloads are evicted prematurely.
Eviction also requires the Merge operation to distinguish between
evicted and non-evicted payloads.
To disambiguate payloads, the metadata table includes generation numbers.
When the switch receives a Split-enabled packet from the NF server, the Merge operation
checks that the generation number in the \PP{} header matches the one in the
metadata table.
If they match, the Merge operation proceeds with adding the payload to the packet.
Otherwise, Merge concludes that the payload was evicted
prematurely and drops the packet and increments the premature eviction counter.

The switch uses L2 forwarding to direct packets to their destination, and
the NF framework processes the packets through
an NF chain.
L2 forwarding and the NF framework run independently of the \PP{} components.

\section{\PP{} Switch Dataplane Design}
\label{sec:design}


\begin{figure}[t]
  \includegraphics[width=\linewidth]{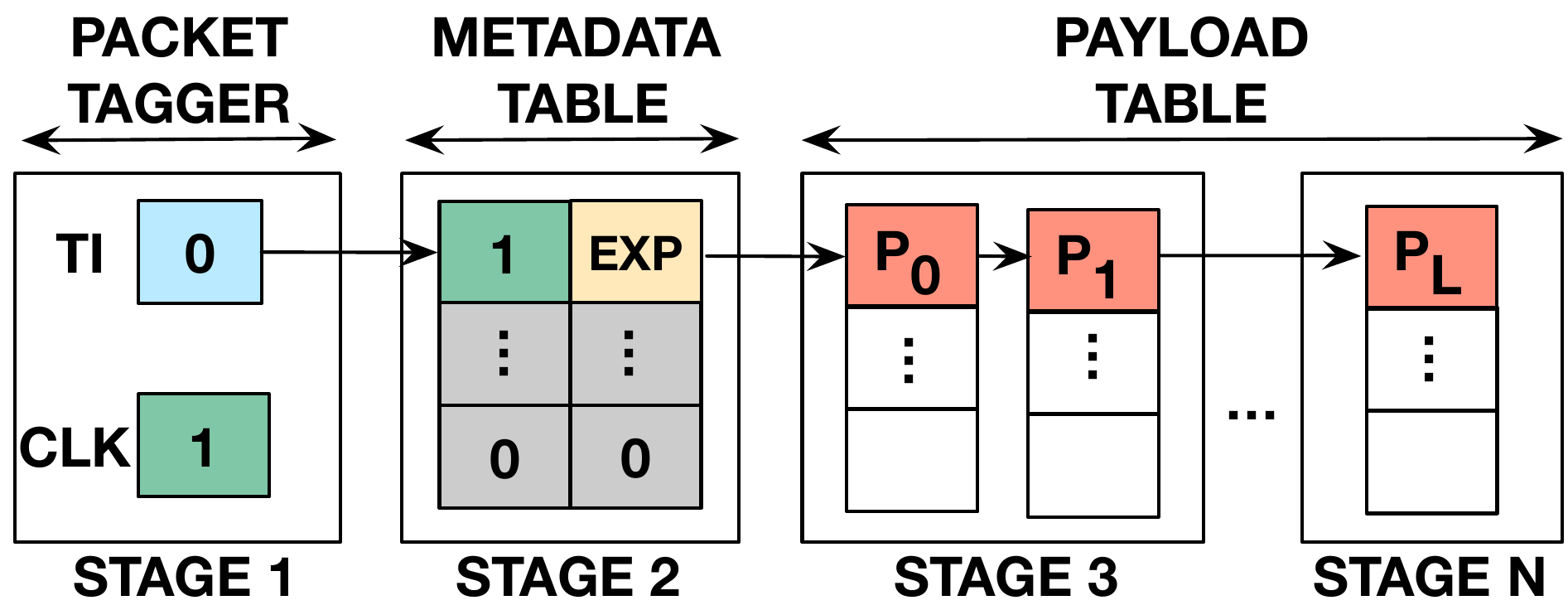}
  \caption{PayloadPark dataplane implementation.
Packet processing proceeds from stage 1 to stage N. Packets are first tagged,
then looked up in the metadata table, and finally stored in the payload table.
The tagger has registers for the table index (TI), which is an index into the
metadata and payload tables, and a clock (CLK).
The metadata table contains two values at each index, the value of the clock when
the index was occupied and an expiry threshold (EXP).
If the index is available for storing payload, its EXP value is 0.
Payload blocks (P\textsubscript{0}, P\textsubscript{1} ... P\textsubscript{L}) are striped across MATs.
}
  \label{fig:SMP_Algo}
\end{figure}


We now map the general purpose design described in
\S\ref{sec:high_level_algorithm} onto the architecture of the
switch dataplane.
Recall that RMT switches process packets by passing them through a series of MATs.

\begin{algorithm}[!tb] \label{VOAlgorithm}
\caption{\textbf{Split operation}}

\hskip0.3em - \textit{M}: Maximum capacity of the lookup table \newline
- \textit{tbl\_idx}: Register for storing table index value \newline
- \textit{clk}: Register for storing the clock counter \newline
- \textit{meta\_tbl[M]}: Register array for storing metadata table \newline
- \textit{pload\_tbl[M]}: MAT-local register array for storing payload block \newline
- \textit{MAX\_EXP}: Pre-configured threshold for evicting payloads \newline
- \textit{MAX\_CLK}: Threshold for resetting the clock \newline
- \textit{SPLIT\_PORT}: Switch port reserved for Split operation \newline
- \textit{hdr.pp}: PayloadPark header \newline
- \textit{meta}: User-defined struct for intermediate results
\begin{algorithmic}[1]
\STATE \textbf{stage} 1:
\STATE \Comment{Packet tagger operations}
    \STATE \doubleindent \textbf{match} in\_port == SPLIT\_PORT:
    \STATE \tripleindent  tbl\_idx = (tbl\_idx + 1) \% M \label{alg1:start_tagger}
    \STATE \tripleindent  meta.tbl\_idx = tbl\_idx
    \STATE \tripleindent clk = (clk + 1) \% MAX\_CLK
    \STATE \tripleindent meta.clk = clk \label{alg1:end_tagger}

\STATE \textbf{stage} 2:
\STATE \Comment{Index into metadata table}
    \STATE \doubleindent \textbf{match} in\_port == SPLIT\_PORT: \tripleindent
    \STATE \tripleindent \textbf{if} {meta\_tbl[meta.tbl\_idx].\PPTTL{} $>=$ 1}:  \label{alg1:start_dec_exp}
        \STATE \fourthindent \Comment{Decrement Expiry threshold}
        \STATE \fourthindent meta\_tbl[meta.tbl\_idx].\PPTTL{} -= 1 \label{alg1:end_dec_exp}
    \STATE \tripleindent \textbf{if} {meta\_tbl[meta.tbl\_idx].\PPTTL{} == 0}: \label{alg1:start_evict}
    \STATE \fourthindent \Comment{tbl\_idx is available for storing payload}
        \STATE \fourthindent meta\_tbl[meta.tbl\_idx].\PPTTL{} = MAX\_\PPTTL{} \label{alg1:start_update_meta_tbl}
        \STATE \fourthindent meta\_tbl[meta.tbl\_idx].clk = meta.clk \label{alg1:end_update_meta_tbl} \label{alg1:end_evict}
        \STATE \fourthindent hdr.pp.is\_enb = 1 \label{alg1:start_set_pp_hdr}
        \STATE \fourthindent hdr.pp.tag.tbl\_idx = meta.tbl\_idx
        \STATE \fourthindent hdr.pp.tag.clk = meta.clk \label{alg1:end_set_pp_hdr}
    \STATE \tripleindent \textbf{else}: \label{alg1:start_turn_off_pp}
        \STATE \fourthindent \Comment{Set all values in \PP{} hdr to zero}
        \STATE \fourthindent hdr.pp = 0 \label{alg1:turn_off_pp}
    \STATE \tripleindent \Comment{Add \PP{} header to the packet}
    \STATE \tripleindent hdr.pp.setValid() \label{alg1:end_turn_off_pp}

\STATE \textbf{stage} 3..N (idx):
\STATE \Comment{Store payload blocks in payload table}
    \STATE \doubleindent \textbf{match} in\_port == SPLIT\_PORT \textbf{and} \label{alg1:start_store_pload}
    \STATE \hskip4.2em hdr.pp.is\_enb == 1:
      \STATE \tripleindent pload\_tbl[hdr.pp.tag.tbl\_idx] = hdr.pload\_block[idx] \label{alg1:end_store_pload}
\end{algorithmic}
\end{algorithm}

\noindent \textbf{Split operation.}  When the switch receives a packet,
it executes the Split operation (Alg. 1).
In stage 1, we generate two subparts of the tag to probe into
the lookup table:
a) an index to find a potential empty location for storing packet payload,
and b) a generation number.
We maintain two counters, a table index (TI) and
a clock counter (CLK),
for indexing into the lookup table.
In stage 1,
we increment both of these counters (Lines~\ref{alg1:start_tagger} -~\ref{alg1:end_tagger})
and roll them over when they
reach their assigned maximum values.
We update the user-defined metadata fields to make the TI and CLK values
available in subsequent stages.

In stage 2, we probe the lookup table to determine if an empty slot
is available.
The TI is an index into the metadata table.
We first read the Expiry value from the slot indicated by the TI.
If it is 0 (meaning that the index is available), we write the clock
value (CLK) and the
Expiry threshold (EXP) into the metadata table at the table index (TI).
For example, in Fig. \ref{fig:SMP_Algo}, assume that when the Split operation started,
the EXP value at element 0 (the TI) was 0, indicating that the  location is available.
We then write CLK and EXP into the zeroth entry of the metadata table.
The Split operation consumes unoccupied locations by assigning the
Expiry threshold and clock value
(Lines~\ref{alg1:start_update_meta_tbl} -~\ref{alg1:end_update_meta_tbl}).
Stage 2 also adds the \PP{} header, sets the Enable bit, and adds tag values to the
header (Lines~\ref{alg1:start_set_pp_hdr} -~\ref{alg1:end_set_pp_hdr}). The \PP{} header fields, including the Enable bit, are set to zero for
cases where the lookup table entry at the table index is occupied (Line~\ref{alg1:turn_off_pp}).

Assuming we found an empty location in stage 2, in stage 3..N,
we store the packet payload.
The payload table is organized as a two dimensional array, where
the columns are spread across MATs.
To match this memory layout,
we break the incoming payload into equal-sized blocks, called \emph{payload blocks}.
The width of a payload block is equal to the width of a single-cell in the 2D array.
Following the example in Fig. \ref{fig:SMP_Algo}, we store the payload at the zeroth row
by striping the payload blocks, $P_0, P_1 ... P_L$, across all the columns
(Lines~\ref{alg1:start_store_pload} -~\ref{alg1:end_store_pload}).
A single stage can execute multiple MATs in parallel, in which case, different
payload blocks are stored in different MATs in the same stage. For brevity,
we show code for a single MAT in each of the 3..N stages in Algorithm 1.

\setlength{\textfloatsep}{0pt}
\begin{algorithm}[!h]
\label{V1Algorithm}
\caption{\textbf{Merge operation}}
\hskip0.3em - \textit{M}: Maximum capacity of the lookup table \newline
- \textit{tbl\_idx}: Register for storing table index value \newline
- \textit{clk}: Register for storing the clock counter \newline
- \textit{meta\_tbl[M]}: Register array for storing metadata table \newline
- \textit{pload\_tbl[M]}: MAT-local register array for storing payload block \newline
- \textit{MERGE\_PORT}: Switch port reserved for Merge operation \newline
- \textit{hdr.pp}: PayloadPark header \newline
- \textit{meta}: User-defined struct for intermediate results
\begin{algorithmic}[1]
\STATE \textbf{stage} 1:
\STATE \Comment{Remove \PP{} header when Enable bit is zero }
    \STATE \doubleindent \textbf{match} in\_port == MERGE\_PORT \textbf{and}
    \STATE \hskip4.2em hdr.pp.is\_enb == 0:
    \STATE \tripleindent hdr.pp.setInvalid() \label{alg2:remove_pp_hdr_stage1}

\STATE \textbf{stage} 2:
\STATE \Comment{Validate Merge requests}
     \STATE \doubleindent \textbf{match} in\_port == MERGE\_PORT \textbf{and} \\
      \hskip4em hdr.pp.isValid() \textbf{and}  hdr.pp.is\_enb == 1:

    \STATE \tripleindent meta.is\_pp\_enb = 0
    \STATE \tripleindent \textbf{if} {meta\_tbl[hdr.pp.tag.tbl\_idx].clk == hdr.pp.tag.clk}:\label{alg2:match_clk}
        \STATE \fourthindent meta.is\_pp\_enb = 1
        \STATE \fourthindent meta\_tbl[hdr.pp.tag.tbl\_idx] = 0 \label{alg2:reclaim_space}
        \STATE \fourthindent meta.tbl\_idx = hdr.pp.tag.tbl\_idx
    \STATE \tripleindent hdr.pp.setInvalid() \label{alg2:remove_pp_hdr}

\STATE \textbf{stage} 3..N (idx):
\STATE \Comment{Read payload blocks from payload table}
\STATE \singleindent \textbf{upon} receiving pkt \textbf{do}
    \STATE \doubleindent \textbf{match} in\_port == MERGE\_PORT \textbf{and} \\ \hskip4em meta.is\_pp\_enb == 1:
    \STATE \tripleindent hdr.pload\_block[idx] = pload\_tbl[meta.tbl\_idx] \label{alg2:start_retr_payload}
    \STATE \tripleindent hdr.pload\_block[idx].setValid()
    \STATE \tripleindent pload\_tbl[meta.tbl\_idx] = 0 \label{alg2:end_retr_payload}

\end{algorithmic}
\end{algorithm}

\noindent \textbf{Payload eviction.}
The Split operation also reclaims memory in the lookup table by cleaning
up long-living payloads.
In the metadata table, we keep track of the Expiry threshold
for every occupied position in the payload table.
If during the Split operation, the TI points to
an occupied location (indicated by non-zero
value of the Expiry threshold), we decrement the Expiry threshold (Lines~\ref{alg1:start_dec_exp} -~\ref{alg1:end_dec_exp} in Alg. 1).
When the associated Expiry threshold reaches zero, we evict the stored payload (Lines~\ref{alg1:start_evict} -~\ref{alg1:end_evict} in Alg. 1)
and reclaim the space for splitting packets.

The value of the Expiry threshold controls how soon payloads are evicted.
An Expiry threshold of 1 indicates that the TI will traverse the
lookup table once before evicting payloads. Higher Expiry thresholds
are more conservative, decreasing the
probability of premature payload evictions, but increasing the time
for which payloads of lost/dropped packets continue to occupy space in the lookup table.
For example, when traffic flows to the NF server but the switch does not
receive any packets back from the NF server, the lookup table will fill
quickly.

With a modest amount of memory, premature evictions will be rare even with an
aggressive Expiry threshold.
Consider a worst-case setup where \PP{}  reserves 2MB of switch memory, stores 160
bytes of payload for every packet, uses an Expiry threshold of 1, and packets arrive at line-rate (40Gbps).
Assuming there is an average 30 $\mu s$ time-delta between Split and Merge operations (including
NF processing), payloads will be prematurely evicted after being stored for approximately 520
$\mu s$, or $17.3\times$ the usual time between Split and Merge operations.
A time-delta of 30 $\mu$s aligns with our observations (\S\ref{sec:perf_real_workload}),
so a maximum time of 520 $\mu s$ gives NFs ample room to return a packet to the switch.

%

The maximum number of lookups for finding empty array slots depends on
switch-specific constraints; in the presented algorithm, we index into the metadata table
once per packet.
In the worst case, the Split operation will not be able to find
a suitable eviction candidate after exhausting the permitted number of lookups.
In such a case, we turn off the Split operation
and set the \PP{} header fields accordingly (Lines~\ref{alg1:start_turn_off_pp} -~\ref{alg1:turn_off_pp} in Alg. 1).

\noindent \textbf{Merge operation.} When the switch receives a packet from
the NF server, we execute the Merge operation (Alg. 2).
In stage 1, we process packets for which \PP{} operation was disabled.
This can happen when the lookup table was full,
and there were no suitable candidates for payload eviction.
%
For such packets, we remove the \PP{} header (Line~\ref{alg2:remove_pp_hdr_stage1}).
No further \PP{} processing is required; we simply forward these packets to their
destination using L2 forwarding.

In stage 2, we validate that the packet's stored payload has not been
evicted by
comparing the
clock values in the \PP{} header and the metadata table (Line~\ref{alg2:match_clk}).
%
If the validation succeeds, we reclaim the space in the metadata
table (Line~\ref{alg2:reclaim_space}) for subsequent Split operations.
%
%
We also remove the \PP{} header (Line~\ref{alg2:remove_pp_hdr}). If the clock values in the \PP{}
header and the metadata table do not match,
we conclude that the payload was prematurely evicted. We drop the packet and
record the drop.  We omit this code for brevity.

In the subsequent stages, we merge the payload back to the validated Merge packets
and remove the stored payload from the lookup table (Lines~\ref{alg2:start_retr_payload} -~\ref{alg2:end_retr_payload}).


\section{Implementation} \label{sec:implementation}

We implemented the \PP{} prototype on 6.4Tbps switch with
Barefoot Tofino ASIC~\cite{tofino_switch}.
The chip has 4 pipes, where each pipe is composed of
Parsers, Match-action pipeline, and Deparsers.
The switch has a total of 64 ports (100Gbps each),
divided into four sets of 16. Each set of 16 share a pipe and its resources.
In our prototype, we reserve ports for \PP{} operation.
The total number of reserved ports is a matter of policy; we change the policy
by changing the configuration in our P4 implementation.
The switch ports that process incoming traffic and the NF server must share
the same pipe, because pipes do not share stateful memory resources.

We implemented \PP{} using approximately 900 lines of $P4_{16}$
code~\cite{p416} that we release in~\cite{pp_code}.
The prototype implements and accesses the lookup table
using P4's register API.
In addition, the packet tagger uses two 2-byte registers for the table index
and the clock counter (see Fig.~\ref{fig:SMP_Algo}).
Thanks to the atomic nature of action execution in P4,
subsequent packets in the match-action pipeline are
guaranteed to get different indexes.
Therefore, each packet has a unique index in the lookup table.
In our design, we can store up to 160 bytes of per-packet payload in the payload table.

We maintain eight counters for monitoring \PP{} operation.
These counters are updated during Merge operation that is described
in Algorithm 2; counters are not shown for brevity.
We include brief explanation of these counters.
In stage 1, we count the number of
packets received from the NF server with Split disabled (i.e., the ENB bit in the
\PP{} header is 0).
In stage 2, we measure the number of Splits, Merges,
and Explicit Drops (discussed in \S\ref{sec:eval_packet_drops}).
In stage 3, we track the total number of evictions and premature payload evictions.
In addition, for disabled Split operations we count: a) how many times the payload
size was less than 160 bytes, and b) how many times the next location
in metadata table was occupied.

\noindent \textbf{Implications of ASIC restrictions.}\footnote{
Due to confidentiality reasons, we omit precise details
of the programmable ASIC.
Instead, we focus on our design approaches to circumvent its restrictions.}
Programmable ASICs (and the Tofino chip) limit the number of
compute and stateful operations per MAT
to guarantee line-rate packet processing.
These limits ensure that \PP{} meets its
design goal of not introducing a latency penalty in NF processing.
However, these constraints have significant implications on \PP{}'s design.

During the Split operation, we do a single lookup in the metadata table.
If that index is occupied, we disable the Split operation.
It is possible that there are empty slots available elsewhere in the
lookup table, but we are not able to find them, because MATs
are restricted to a single register operation per packet.
Laying out stateful memory
as arrays, or more accurately, a circular buffer, neatly coexists with this
restriction.
Usually, Split and Merge process packets in the same (FIFO) order.
Thus, if we allocate space in the metadata table sequentially, as the
table index works its way through the array,
Merge operations reclaim memory at earlier positions in the array.
By the time the table index wraps around,
it should find empty spots. This access pattern optimizes \PP{} operation for
the common case (FIFO order). In the worst case, where long-living payloads are occupying the lookup table,
we disable the Split operation.

We apply the Split operation only when the payload length exceeds the number of
per-packet bytes that we can store (160 bytes in our implementation).
This decision prevents wasting memory resources.
Every stored payload reserves a table index and to fully utilize per-index memory, the payload size must be at least 160 bytes.
Turning off Split operation for payloads smaller than 160 bytes prevents this wastage.

\noindent \textbf{NF framework integration.}
Our \PP{} prototype does not require changes to shallow NFs.
The \PP{} header replaces part of the payload, which remains unexamined by shallow NFs.
We use switch port numbers to disambiguate between Split and Merge operations, so
the NF server does not need to change the Merge opcode.

\section{Evaluation}
\label{sec:eval}

\begin{figure}[t]
\centering
   \includegraphics[width=1\linewidth]{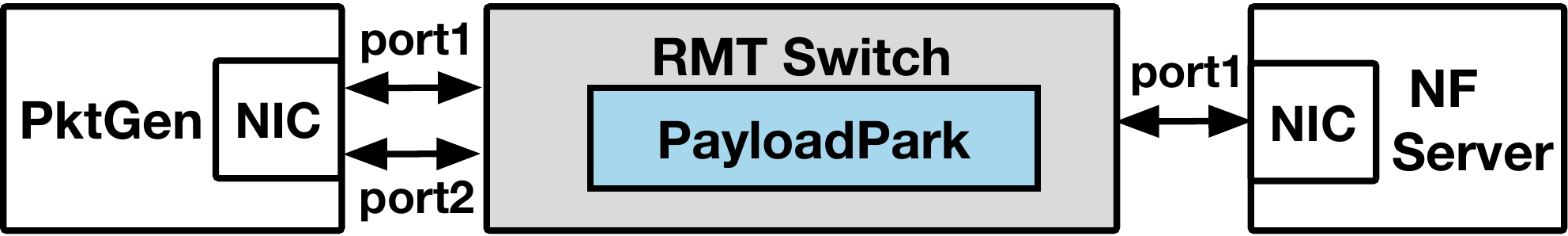}
\caption{Experimental setup.}
\label{fig:ExpSetup}
\end{figure}

The evaluation answers the following research questions:
\newline \textbf{Performance.} Does the prototype satisfy the performance
goals of providing goodput improvement without hurting latency?
What is the performance profile of NFs that benefit from \PP{}?
\newline \textbf{Operation with limited resources.}
Does the prototype use the switch resources efficiently?
We evaluate the effects of parameters, such as Expiry threshold and
percentage of memory reserved, on overall performance.

\subsection{Methodology}
\label{sub:methodology}

\noindent \textbf{Setup.} We deployed \PP{} on a
6.4Tbps switch with Barefoot Tofino ASIC~\cite{tofino_switch}.
For the non-\PP{} setup (our baseline), the switch uses L2 forwarding to pass
traffic between the traffic generator and the NF server.

We use two NF frameworks, OpenNetVM~\cite{opennetvm} and NetBricks~\cite{NetBricks},
to evaluate our prototype. OpenNetVM is built on top of DPDK and
runs NF chains in Docker containers~\cite{opennetvm}.
NetBricks is a DPDK-based framework written in Rust, but
does not use containers to isolate memory between NFs.
We evaluated \PP{} using two dual port NICs: an Intel 82599ES 10 GE NIC and an Intel XL710 40 GE NIC.

We use PktGen, a DPDK-based traffic generator to saturate the NF server with UDP packets.
We run PktGen on a dual NUMA node, 2.4 GHz Intel Xeon E5-2407 v2 server with 8 CPU cores
and 48 GB RAM.
Fig.~\ref{fig:ExpSetup} shows the experimental setup.
We connect two ports of the traffic generator's NIC to the switch to saturate the NF server.
One port is not sufficient, because \PP{} reduces
data that the switch transmits to the NF server.
The NF server is connected to the switch using a single port on the NIC.
We use identical NICs for the traffic generator and the NF server.

The NF server is a four NUMA node, 60 core machine with the 2.3 GHz Intel Xeon E7-4870 v2 processor
and 512 GB RAM. With OpenNetVM, we reserve 3 cores for the OpenNetVM manager, and
each NF in the chain is pinned to one core. With NetBricks, we use 4 cores to run the NF
chain. We can scale up the NF framework by using additional cores
or ports, but this tuning is orthogonal to our evaluation,
because  \PP{} is an in-network
optimization.
We reserve 8 GB of memory backed by hugepages on each NUMA node.
We use two NF chains, $Firewall \rightarrow NAT \rightarrow LB$  and $Firewall \rightarrow NAT$, to evaluate our
prototype. The firewall linearly probes through a list of blocked IP addresses. The firewall in the three-NF chain
has 20 rules, and the two-NF chain has a single rule in its firewall.
The load balancer is based on the Maglev load-balancer~\cite{Maglev}.
The NAT is based on MazuNAT from NetBricks~\cite{NetBricks}.

\noindent \textbf{Evaluation metrics.} \PP{} is a \emph{goodput}
optimization, which
we measure from the RMT switch's perspective.
We use the packet header as the unit of useful information.
In our evaluation, throughput of 10 Mpps corresponds to 3.36 Gbps of goodput
in the baseline and \PP{},
because the packet header (including Ethernet, IPv4 and UDP header)
length is 42 bytes (336 bits).
We also measured average end-to-end packet processing latency
from and to the traffic generator
to validate our performance goals (\S\ref{sec:design_goals}).
We also measured PCIe bandwidth utilization on the NF server using Intel's Processor Counter Monitor~\cite{pcm}.
We consider the system to be healthy when the packet drop rate is
below 0.1\%; we use this threshold to measure peak goodput of
\PP{} and the baseline.
Unless mentioned explicitly, all evaluations have no premature payload
evictions -- a prerequisite for functional equivalence.

\begin{figure}[t]
  \includegraphics[width=\linewidth]{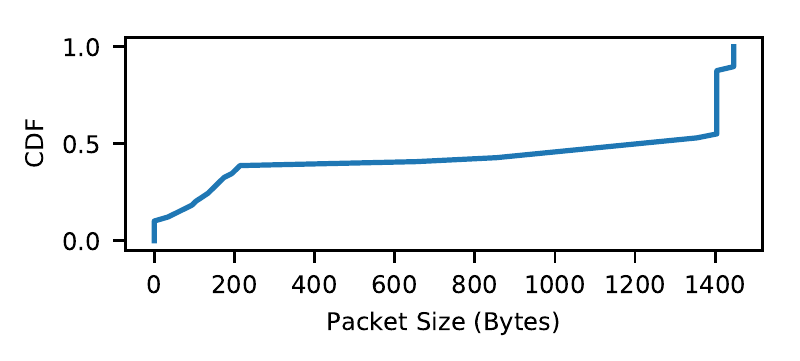}
  \caption{Packet size distribution for simulating enterprise datacenter traffic pattern.}
  \label{fig:PktSize_CDF}
\end{figure}

\noindent \textbf{Workloads.}
We replay PCAP files to simulate an enterprise datacenter traffic pattern.
Our PCAP (shown in Fig.~\ref{fig:PktSize_CDF}) reproduces the packet-size
distribution of enterprise datacenters reported by Benson et al.~\cite{dctraffic}.
The packet sizes have a bimodal distribution with an average packet size of 882 bytes.
Recall that we do not split packets whose payloads have fewer than 160 bytes.
In this workload, 30\% of the packets have fewer than 160 bytes for which
we add the \PP{} header and set the ENB bit to zero.
We collect performance metrics by replaying UDP packets over a period of 2 minutes.
For all experiments, we report the average of three runs.
We also evaluate our prototype with fixed-sized packets to determine
the range of packet sizes that benefit from the \PP{} deployment.

\noindent \textbf{Experiments.}
For macro-benchmarks, \PP{} reserves about 26\% of switch memory,
and we set the Expiry threshold to 1.
We evaluate \PP{} on our 40 GE NIC using both the datacenter traffic pattern (Fig.~\ref{fig:PktSize_CDF})
and fixed packet sizes for single Firewall and NAT NFs and the NF chain
consisting of Firewall $\rightarrow$ NAT.
With the 10 GE NIC, we evaluate the
goodput gain with Firewall $\rightarrow$ NAT $\rightarrow$ LB chain.
We also used packet recirculation to increase the number of stored payload bytes
and evaluated the Firewall $\rightarrow$ NAT $\rightarrow$ LB chain
using the NetBricks framework with a 10GE NIC.

\begin{figure}[t]
\centering
 \includegraphics[width=1\linewidth]{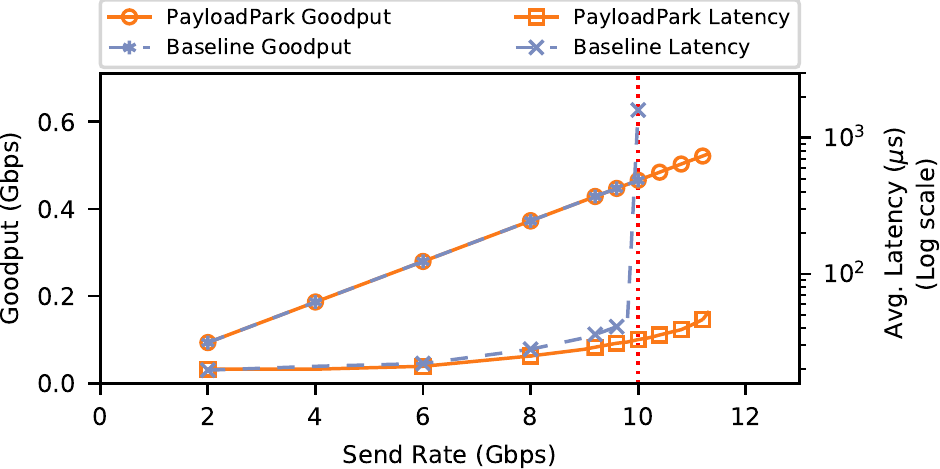}
  \caption{ Results for $FW \rightarrow NAT \rightarrow LB$. The vertical
  red line at X=10 Gbps highlights baseline link capacity. For X$<$10 Gbps,
  the maximum difference between peak and average latency is 0.66 $\mu$s
  for baseline and \PP{}. At X=10 Gbps, this difference is 21 $\mu$s
  and 0.66 $\mu$s for baseline and \PP{}, respectively.}
\label{fig:PeakGoodput}
\end{figure}

We evaluate \PP{} when multiple NF servers share the same switch.
Each server runs a MAC address swapper.
The switch has 4 pipes, and each pipe is connected to two NF servers.
Each NF server processes traffic from its dedicated traffic-generator.
We use the previously described 8 core servers for both the traffic generator
and NF servers.
We increase the reserved memory resources on the switch to about 40\%,
and slice the reserved memory
amongst NF servers sharing the same pipe.
We use 384 byte packets, because
for a fixed link capacity, smaller packets
produce a higher packet rate that puts more memory pressure on the switch.

For micro-benchmarks,
we vary implementation parameters, such as
the reserved memory on the Tofino ASIC and Expiry thresholds.
We evaluate \PP{} using NFs with different computational costs.
To create NFs of varying computational cost,
we take a MAC address swapper and add a busy loop.
We measure per-packet CPU cycles using the RDTSC counter~\cite{rdtsc}.

\subsection{Macro-Benchmark Results}

\subsubsection{Performance Improvement}
\label{sec:perf_real_workload}

Fig.~\ref{fig:PeakGoodput}
shows the goodput and latency improvement of \PP{} with the NetBricks framework.
We used the $FW \rightarrow NAT \rightarrow LB$
chain with the 10 GE NIC.
We observed similar results with OpenNetVM, omitting the results for brevity.
The experiment shows that
\PP{} can process more traffic than the baseline without a latency
penalty.
We also evaluated the $FW \rightarrow NAT$ chain with a 40 GE NIC using OpenNetVM framework.
With the 40 GE NIC, \textbf{we observe a 15.6\% goodput improvement and no latency penalty}.
We did not make any code changes to either NF framework, demonstrating that
\PP{} easily integrates with NF frameworks.

Relative to the baseline,
\PP{} reaches the latency cliff at a higher packet send rate.
In the \nonPP{} setup, the average latency increases sharply as
the network link approaches saturation, while
\PP{} has no such spike, because
the switch-to-NF server link does not approach saturation.

The bimodal packet distribution is representative of datacenter traffic,
but it yields modest goodput gain with \PP{}.
The reason is two-fold:
1) 30\% of the traffic is not Split because packet size is less than 160 bytes,
and 2) the average packet size is 882 bytes, so we don't truncate a substantial
proportion of the payload.

\begin{figure}[t]
  \includegraphics[width=\linewidth]{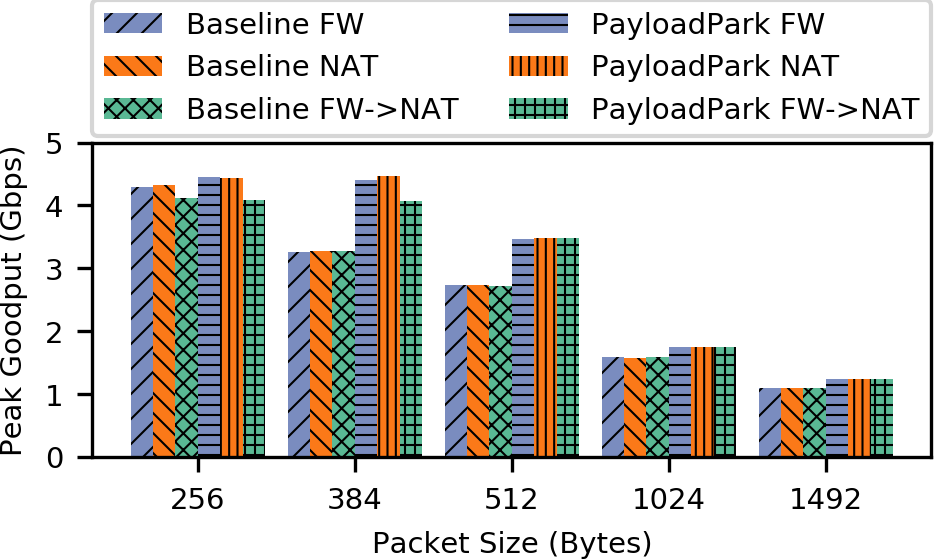}
  \caption{(Higher is better) Goodput with different packet sizes.}
  \label{fig:NFvsPktSize}
\end{figure}

We also found that
\textbf{\PP{} reduces PCIe bus load by 12\% on the NF server at all send rates}
(the graph is excluded for brevity).
Assuming that future generations of RMT switches have more memory,
the PCIe bandwidth savings will increase, because we can store more payload
bytes on the switch.

An alternative deployment option is to deploy \PP{} on a SmartNIC.
However, this approach will not improve goodput on the link.
Additionally, prior work showed that NFs can run directly on
SmartNICs~\cite{clickNP, E3}, in which case, we could derive the dual benefit
of a performance gain from the SmartNIC and goodput gains from
\PP{} on the switch.

\subsubsection{Goodput Improvement with Fixed Packet Sizes}
\label{sec:diff_nfs_diff_pkt_size}

\begin{figure}[t]
  \includegraphics[width=\linewidth]{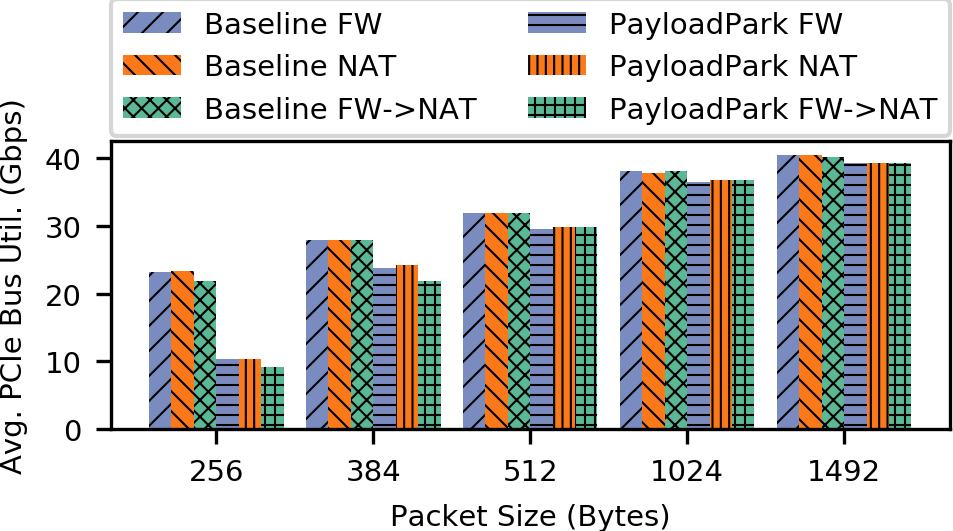}
  \caption{ (Lower is better) PCIe bus utilization with different packet sizes.}
  \label{fig:NFvsPktSize_PCIe}
\end{figure}

Fig.~\ref{fig:NFvsPktSize} and Fig.~\ref{fig:NFvsPktSize_PCIe}  show \PP{}'s behavior with
different packet sizes and NF chains using the 40 GE NIC and OpenNetVM framework.
\textbf{The goodput improvement between \nonPP{} and \PP{} is 10-36\% for 384 to 1492 byte packets.}
We see a larger goodput gain at smaller packet sizes (384 and 512 byte packets), because we truncate a larger proportion of each packet for small packet sizes.
Also, the $FW \rightarrow NAT$ chain has lower goodput gain than individual NFs,
because
the NF server does more per-packet computation, making OpenNetVM compute bound sooner.
Similarly, for 256 byte packets, the goodput gain is negligible, because for a fixed bandwidth,
smaller packet sizes put more compute pressure on the NF server and more memory pressure on the switch than large packet sizes.
But, \textbf{\PP{} reduces PCIe bus load by 58\% for 256 byte packets (Fig.~\ref{fig:NFvsPktSize_PCIe})};
the reduction is proportional to the number of payload bytes stored in the switch.

Effective PCIe bandwidth decreases when we transmit small packets.
Neugebauer et al. show that although the effective PCIe Gen 3 x8 interface
(that we use in our experiments) bandwidth is expected to accommodate
40 GE link for all packet sizes, current NIC deployments do not match
the expectation (see Fig. 1 in pcie-bench paper~\cite{pcie_bench}).
For example, a modern NIC with DPDK driver (that we use) cannot operate
at 40 Gbps for packets smaller than 170 bytes; PCIe becomes bottleneck.

\PP{} might seem to exacerbate PCIe congestion as it reduces the packet sizes
by parking (the part of) the payload in the switch.
However, \PP{} still offers a net goodput gain,
despite the lower effective PCIe bandwidth, because of two reasons.
First, the ratio by which \PP{} improves the goodput is always higher than
the effective PCIe bandwidth reduction ratio. For example, when all packets are
256 bytes, 40 GE link delivers 6.5 Gbps goodput
(for 42 byte header). In our implementation, \PP{} will park 160 bytes
and send 103 byte packet (96 byte payload + 7 byte \PP{} header)
over the PCIe bus that can offer only 26 Gbps effective bandwidth
(instead of 40 Gbps with 256 byte packets). 26 Gbps accommodates
31 million 103 byte packets that gives 10.4 Gbps goodput.
Thus, the goodput with \PP{} (10.4 Gbps) is higher than the baseline
(6.5 Gbps) despite PCIe bandwidth reduction.
Second, the datacenter traffic characteristics that we adopt from
Benson et al.~\cite{dctraffic} and other studies~\cite{caida-dataset}
have average packet size of 800 bytes -- large enough to mask PCIe bandwidth
reduction and make it less of a concern for \PP{}'s goodput optimization.

\begin{figure}[t]
  \includegraphics[width=\linewidth]{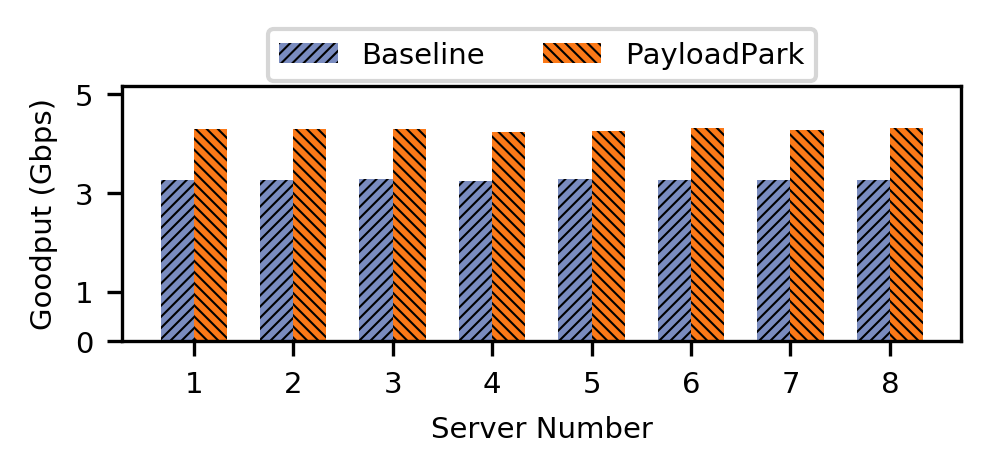}
  \caption{(Higher is better) Goodput with 384 byte packets for 8 NF servers numbered from 1-8 on x-axis.}
  \label{fig:ScalabilityTput}
\end{figure}

\begin{figure}
  \includegraphics[width=\linewidth]{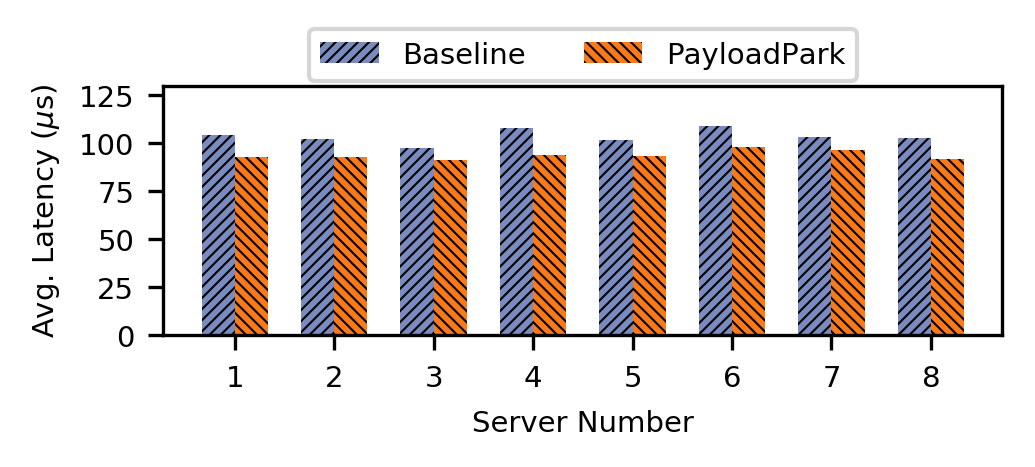}
  \caption{ (Lower is better) Latency with 384 byte packets for 8 NF servers numbered from 1-8 on x-axis.}
  \label{fig:ScalabilityLatency}
\end{figure}

\subsubsection{Multiple NF Server Setup}

We next examine how \PP{} can benefit multiple NF servers, since
performance isolation is important in multi-tenant clouds.
We simulate such a setup by connecting the switch to 8 NF servers.
Fig.~\ref{fig:ScalabilityTput}  and Fig.~\ref{fig:ScalabilityLatency}
show that \textbf{all 8 NF servers exhibit consistent performance improvement
with an average goodput gain of 31.22\% and latency win of 9.4\%.}
These latency savings are on the PCIe bus,
because \PP{} copies less data between the NIC and CPU for each packet.
This experiment also shows that \PP{} efficiently uses the on-chip memory
resources, because it can service traffic for multiple NF servers.
Static slicing of switch resources between NF servers ensures
performance isolation
in the presence of heavy-hitting traffic from a subset of customers.
This also protects individual customer's payloads from being evicted by other
customer's traffic flowing through the same switch.
The goodput gain of the server is a function of memory reserved
for the operation of that server.
We discuss the impact of varying the reserved memory
in \S\ref{sec:memory_bounds}.

Instead of static slicing, there is potential to further improve \PP{}'s memory
efficiency by dynamically reallocating memory between different NF servers to
match their workload.
We leave this optimization as future work.

\begin{figure}[t]
  \includegraphics[width=\linewidth]{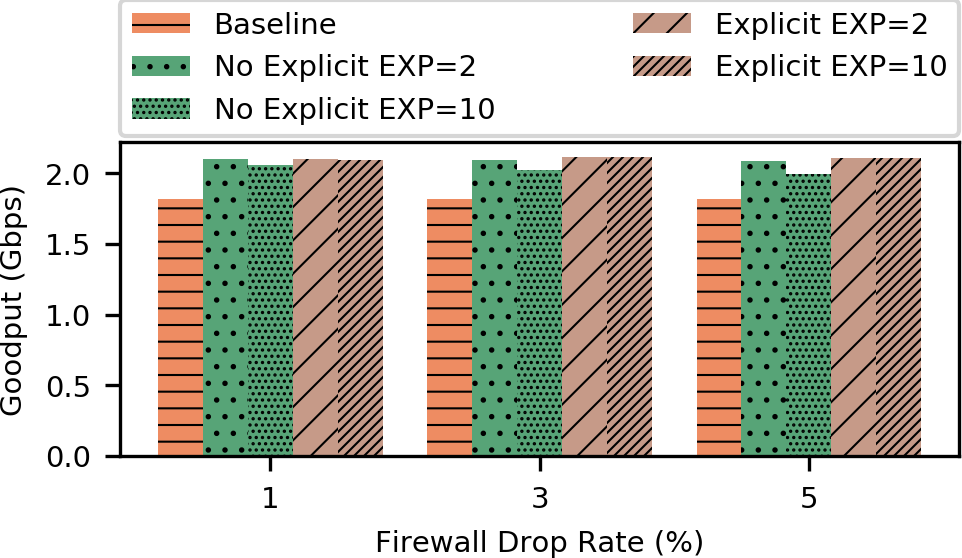}
  \caption{(Higher is better) Goodput using Explicit and No Explicit Drop packets for $Firewall \rightarrow NAT$.
  EXP=2 and EXP=10 labels denote Expiry threshold of 2 and 10 respectively.
  Standard error in goodput measurements is less than 0.26Mbps.
  The graph shows that \PP{} has better goodput than baseline.}
  \label{fig:Explicit_vs_ImplicitDrop}
\end{figure}

\subsubsection{Payload Eviction and Explicit Payload Drops}
\label{sec:eval_packet_drops}
In addition to implementing payload eviction in the switch dataplane, we
explored the option of making small code changes in the NF framework to
explicitly notify the switch when the NF decides to drop a packet.
Explicit Drop notifications provide the ground truth to evaluate the payload
eviction policy, because the NF notifies the switch as soon as a payload can be
evicted.

We added 50 lines of code to the NF framework
(OpenNetVM~\cite{opennetvm}) to send \textit{Explicit Drop} packets to the
switch.
The NF framework marks an incoming packet as dropped by changing the opcode
("OP" bit in Fig.~\ref{fig:SMP_Header}), truncating the packet payload,
and sending the resulting packet back to the switch.
Drop is a special case of Merge that just reclaims memory after validating the
tag.
Explicit drops are an optional optimization; the payload evictor already reclaims memory on the
switch.

Fig.~\ref{fig:Explicit_vs_ImplicitDrop} compares the goodput with and without
explicit drops to different expiry thresholds.
We use the workload shown in Fig.~\ref{fig:PktSize_CDF} for the $Firewall
\rightarrow NAT$ chain.
The firewall blocks packets using a single rule in its
Access Control List, and we vary the proportion of blocked IP addresses
to control the drop rate at the firewall.
Unlike the firewall benchmark recommendations~\cite{FW_RFC}, we do consider
dropped packets in our measurement since we measure goodput from the RMT switch
perspective (see \S\ref{sub:methodology}).

Fig.~\ref{fig:Explicit_vs_ImplicitDrop} shows that aggressive eviction policy
(EXP=2) performs comparably to Explicit Drop notifications.
With a more conservative eviction policy (EXP=10), goodput drops because more
dropped packets occupy space in the lookup table.
Overall, the Expiry threshold presents a trade-off between effective memory
utilization and protection against premature payload evictions.
Explicit Drops \emph{in combination} with conservative payload eviction balance this trade-off.
Fig.~\ref{fig:Explicit_vs_ImplicitDrop} shows that a conservative eviction
policy when combined with Explicit Drops (Explicit EXP=10)
performs comparably to an aggressive eviction policy (No Explicit EXP=2).
The benefit of Explicit Drops comes at the cost of small code changes to the
NF framework, but this cost is a one-time investment.

\subsubsection{Effect of Packet Recirculation}

\begin{figure}[t]
  \includegraphics[width=\linewidth]{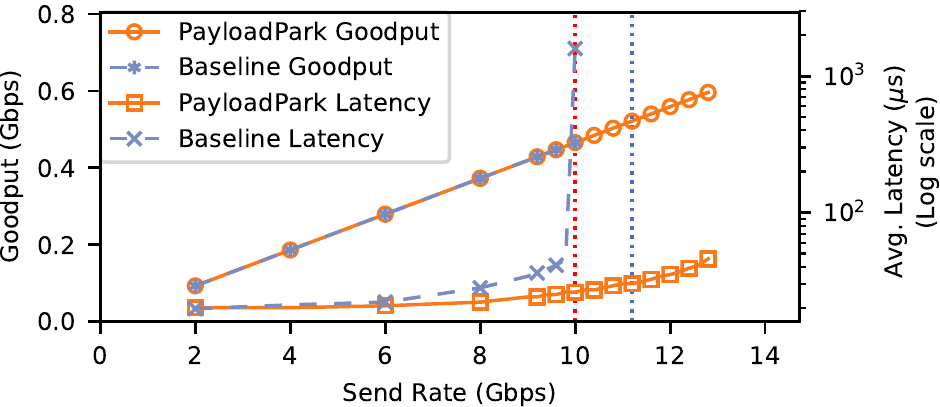}
  \caption{Goodput and latency improvements of PayloadPark with
  $FW \rightarrow NAT \rightarrow LB$ chain on NetBricks.
  This is the recirculation-enabled version of the experiment described in
  Fig.~\ref{fig:PeakGoodput}.}
  \label{fig:Recirc_results}
\end{figure}

In our prototype, we increase the number of stored payload blocks by recirculating
packets in the packet processing pipeline. Recall that
we store payload blocks by striping them across stages 3 to N of a single pipe
(see Fig.~\ref{fig:SMP_Algo}).
Using packet recirculation, we stripe payloads in all the stages of a second pipe
in addition to payload blocks stored in the first pipe. \textbf{Recirculation increases the stored payload size from 160 bytes to 352 bytes.}

Fig.~\ref{fig:Recirc_results} shows goodput and latency with $FW \rightarrow NAT \rightarrow LB$
using the NetBricks framework with the 10 GE NIC. The vertical red line
(at X=10 Gbps) and blue line (at X$\approx$11 Gbps) highlight the peak send rate
that the baseline and our prototype can sustain (without recirculation).
We observe a 28\% goodput improvement --
approximately twice that of
the prototype without recirculation. A  single packet recirculation
induces a latency penalty on the order of 10s of ns~\cite{dejavu}. But, \textbf{we do not observe any
end-to-end
latency penalty thanks to the reduced PCIe latency caused by the additional payload bytes
stored in the switch.}  We also observe a 23\% reduction in PCIe bus load
for all send rates before the baseline link gets saturated.
With these results,
we conclude that goodput improvement and PCIe bandwidth savings increase with an
increase in the stored payload size.

\subsubsection{Functional Equivalence}

We validated functional equivalence by comparing the packets received at the
traffic generator upon return from NF chains in the \PP{} and
baseline deployments.
We used  a single NF that swaps MAC addresses for our evaluation.
We used DPDK-pdump to sniff packets at the traffic generator's NIC.
The resulting PCAP files are identical
and switch metrics report no premature payload evictions,
which gives us confidence that
PayloadPark and baseline deployments are functionally equivalent.

\subsection{Micro-Benchmark Results}
\label{sec:microbenchmarks}

\subsubsection{Impact of Reserved Memory}
\label{sec:memory_bounds}

\begin{figure}[t]
  \includegraphics[width=\linewidth]{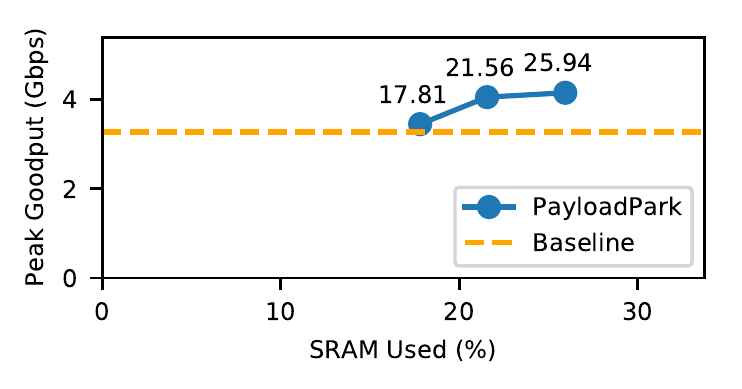}
  \caption{(Higher is better) Goodput compared to percentage of reserved
  memory on the Tofino chip. Labels on points show specific memory usage.}
  \label{fig:MemoryUtil}
\end{figure}

The amount of memory that \PP{} reserves on the switch presents a
trade-off between goodput gain and memory available for implementing additional
P4 functionality.
We use 384 byte packets with the $Firewall \rightarrow NAT$ chain to stress the
memory resources at the switch.
We set the Expiry threshold to 1 and increase the traffic rate until \PP{} begins
to evict packets prematurely.
The number of premature payload evictions is an important metric that determines
functional equivalence of the \PP{} deployment. To achieve functional
equivalence, there must be zero premature payload evictions.
We test with EXP=1, because if there are no premature payload evictions with an
aggressive payload eviction policy (EXP=1), the system will be functionally
equivalent with more conservative (EXP $>$ 1) payload eviction policies.

Fig. \ref{fig:MemoryUtil} shows the peak traffic send rate that exhibits no
premature payload evictions as we increase the percentage of memory allocated for \PP{}.
\emph{First, we see that the \PP{} optimization provides goodput gain with less than
26\% Tofino chip memory.}
Additionally, with a little more than 17\% of switch memory resources, the prototype can sustain
at most 3.44 Gbps goodput. Beyond this rate, the Expiry threshold is not
high enough to prevent premature evictions.
For example, at an incoming
traffic goodput rate of 3.55 Gbps (send rate of 32.45 Gbps), we observed that 0.03\% of incoming payloads
are being prematurely evicted (not shown).
Cloud-providers should not use \PP{} when co-located P4 artifacts
are memory-intensive, and there is insufficient memory for \PP{} operation.

\subsubsection{Resource Utilization}
\PP{} is designed to take advantage of the spare capacity already available in Tofino switches.
Table~\ref{table:ResourceUtilization} shows the switch  resources used by
the \PP{} prototype (excluding L2 forwarding).
Despite being memory-intensive, our average per-stage SRAM utilization is 25.94\%, and it
is comparable to prior work (SilkRoad~\cite{Miao2017SilkRoad}, BurstRadar~\cite{BurstRadar}).
\PP{}'s SRAM utilization is not uniform across all stages in the Match Action Unit;
the peak per-stage memory utilization is 33.75\%.
This memory is sufficient for supporting 4 NF servers (one on each pipe).
Our peak memory utilization is 48.75\% for 8 NF servers,

\begin{table}[t]
\begin{tabular}{ ccc }
\toprule 
\textbf{Resource Name} & \textbf{\makecell{PayloadPark\\ Percentage  Res. Util.}} \\
\midrule 
\emph{\makecell{SRAM  (4 NF servers)}}  & \makecell{25.94\% (Avg.) / 33.75\% (Peak) }  \\
\hline
\emph{SRAM  (8 NF servers)} &  \makecell{38.23\% (Avg.) / 48.75\%  (Peak)}\\
\hline
\emph{TCAM} & 0.69\%   \\
 \hline
\emph{VLIW} & 14.58\%  \\
 \hline
\emph{\makecell{Exact Match \\Crossbar}} & 16.47\%   \\
 \hline
\emph{\makecell{Ternary Match \\ Crossbar}} &0.88\%    \\
 \hline
\emph{\makecell{Packet Header \\ Vector}} &37.65\%    \\
\bottomrule 
\end{tabular}
\caption{Resource Utilization on the Tofino chip.}
\label{table:ResourceUtilization}

\end{table}

\PP{} efficiently uses additional hardware resources such as PHV and VLIW (discussed in \S\ref{sec:background}).
We use only 37.65\% of the PHV resources, which is comparable to our overall
memory consumption and therefore, not a limiting resource.
The other resources that \PP{} uses, including VLIW, TCAM, and
crossbars also have less than 20\% utilization.
Overall, our implementation efficiently uses on-chip memory resources
and leaves sufficient resources for implementing additional P4 functionality.


\subsubsection{Impact of NF CPU Cycles}
 \label{sec:busyCPUNFs}

\begin{figure}[t]
  \includegraphics[width=\linewidth]{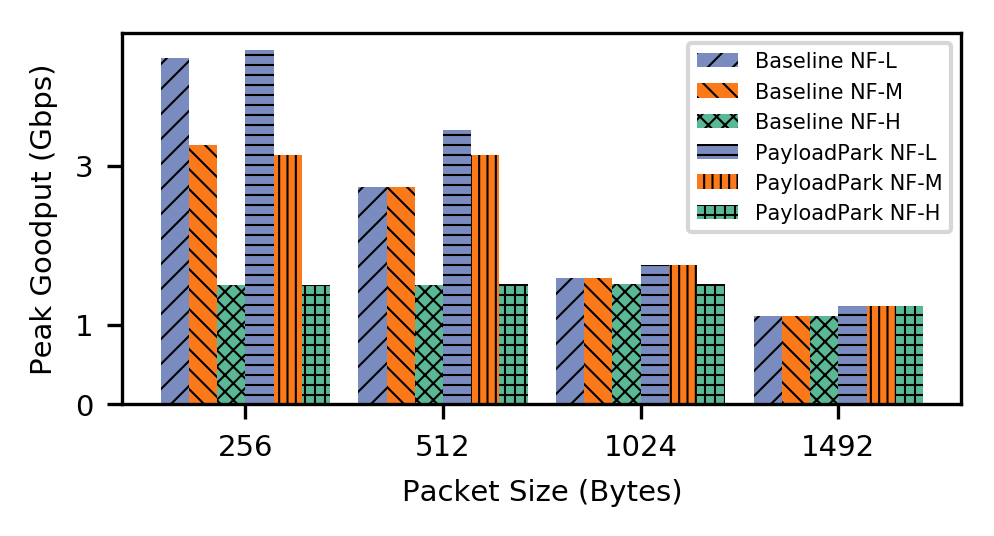}
  \caption{(Higher is better) Goodput with different NFs: NF-Light, NF-Medium and NF-Heavy.}
  \label{fig:BusyCPUCycles}
\end{figure}

We now examine the impact of NF packet processing
time on the achievable peak goodput of the \PP{} prototype.
Fig. \ref{fig:BusyCPUCycles} shows the goodput gain for
4 different packet sizes and 3 different
NF computation loads, NF-Light,
NF-Medium, and NF-Heavy, with approximately 50, 300, and 570
average CPU cycles per packet, respectively.

With 1492 byte packets, \PP{} consistently yields better goodput than the
baseline, because at large packet size (and smaller packet send rate)
OpenNetVM does not become compute bound.
However, for smaller packet sizes ($<= 1024$ bytes), we do not achieve goodput
gain with NF-Heavy, because OpenNetVM
becomes compute bound when the send rate exceeds 5 Mpps (goodput of 1.68 Gbps).
Sending additional traffic
results in packet drops at the NF server NIC in both \PP{} and baseline deployments.

For 256 byte packets, we see negligible (2.1\%) goodput improvement for
NF-Light. 
For small packet sizes, the traffic generator has to
send more packets to sustain the requested transmit rate.
This puts more memory pressure on the switch, since we store a fixed
number of bytes per packet. However,
NF-Light is quick enough to free switch memory, thereby avoiding
premature payload evictions.
For NF-Medium, the \nonPP{} has 3.87\% better goodput than \PP{},
because the increased NF computation increases NF latency resulting in
premature payload evictions.
These results and those presented in \S\ref{sec:diff_nfs_diff_pkt_size}
indicate that larger packet sizes ($>$= 384 bytes)
show better goodput improvement than 256-byte packets.
We can better utilize switch memory by
increasing the minimum payload size threshold of 160 bytes to 384 bytes.

\subsubsection{Impact of \PP{} on NF Frameworks}

\begin{figure}[t]
  \includegraphics[width=\linewidth]{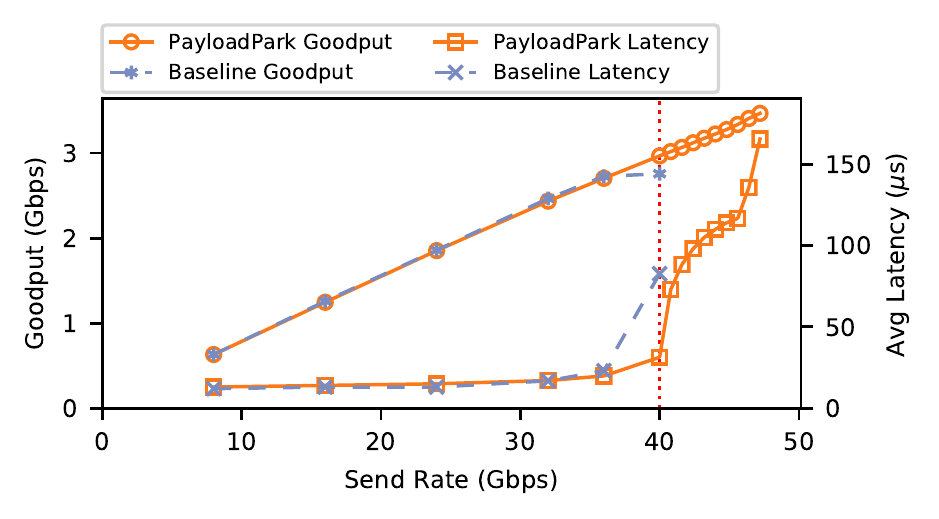}
  \caption{Goodput and latency of PayloadPark and Baseline
  with 512 byte packets and $Firewall \rightarrow NAT$ chain.
  The vertical red line at X=40 Gbps highlights baseline link capacity.}
  \label{fig:LatencyTputTrend}
\end{figure}

Overall, section \ref{sec:perf_real_workload} showed that \PP{} improved latency
with a variety of different packet sizes.
In that case, \PP{}'s latency is better, because large packets dominate the
workload with an average packet size of 882 bytes.
However, for smaller packet sizes, since \PP{} causes the NF server to process
more packets per second, it adds compute strain on the server and latency can
spike.
To illustrate this effect,
Fig.~\ref{fig:LatencyTputTrend} shows goodput and average
latency for the $Firewall \rightarrow NAT$ chain running using OpenNetVM with 512 byte packets.
As shown, \PP{} continues to process packets beyond 33.6 Gbps,
while the \nonPP{} is capped at 33.6 Gbps.
Before the baseline saturates, \PP{} results in lower average latency than the baseline,
because \PP{} reduces per-packet bandwidth consumption between the switch and
the NF server.
But, for both baseline and \PP{}, latency increases once the send rate exceeds
33.6 Gbps.
This latency increase is not a fundamental shortcoming of \PP{}.
Optimizations in the NF framework are orthogonal to our work, and this
latency increase can be addressed by using \PP{} with an NF framework that
maintains line rate at small packet sizes.
It is important to note that this behavior only occurs with small, fixed-size packets.
Overall, \PP{} can improve goodput without latency penalty when packet sizes follow the
enterprise traffic pattern.

\section{Discussion}

Overall, our \PP{} prototype improves goodput on the link between the NF
server and the switch for all tested packet sizes, while working within the
limitations of the RMT switches, and without impacting latency. These results
show that \PP{} is a promising optimization that is already effective on
today's RMT switches.

\noindent \textbf{Limitations.}
\PP{} increases
packet rate on the links, but the overall performance gain is capped by the
performance profile of the end-host NF server. If the NF server setup cannot process the incoming
packet rate in the baseline setup, the overall setup will not benefit from \PP{},
because additional packets that \PP{} transfers over the link will
get dropped at the NF server.
NF frameworks that use commodity hardware are compute bound and this limits the
goodput gain. This is reflected in our choice
of simple NFs and shorter NFs for evaluation with the 40 GE NIC.
With the 10 GE NIC, goodput gain is capped by the NIC capacity.
Also, longer packet processing latency (longer NF chain or
slower NF framework) increase the memory pressure on the switch, limiting
goodput gain with \PP{}. In the worst case, when switch memory is exhausted, \PP{}
gracefully stops parking payload bytes on the switch and adds a fixed
\PP{} header overhead (of 7 bytes) per packet.

\noindent \textbf{Scalability across switches.}
\PP{} is a general optimization that can be deployed on any switch in datacenter topology: core,
aggregate, and the ToR switch.
In our prototype, goodput gain is achieved
by storing payload in a ToR switch. We can further increase the goodput gain, and
distribute memory pressure by striping the
packet payload across multiple switches in the packet path.
When a packet arrives at the cloud-provider's infrastructure,
all switches can perform Split and Merge.

\noindent \textbf{Decoupling boundary.}
In our prototype, we used the UDP header as the header-payload decoupling boundary.
We can trivially change this boundary if the NF server needs to inspect more or fewer fields. 
For example, if the NF chain
reads only the first $N$ bytes of the payload, we can store the rest of the payload in the programmable ASIC memory.
For example,
Fernandes et al. propose Slim-DPI that analyzes a fraction of the payload
to classify packets~\cite{Fernandes2009LWDPI}. \PP{} can be applied to
Slim-DPI-like NF chains.
The decoupling boundary can also be dictated by the protocol, such as TCP.
Our current prototype works with all protocols, but if the NF server needs to
inspect protocol-specific fields beyond TCP/UDP headers, the boundary can be extended to include
these bits in the header as needed.
We leave investigating the impact of different decoupling boundaries to future work.

\noindent \textbf{Failure scenarios.}  The PayloadPark deployment and the
baseline setup deal with three failure scenarios:
\newline 
1) Link failures:
Both deployments are equally susceptible to link failures and accidental
disconnects. All in-flight  messages in disconnected link(s) will be lost.
Incoming traffic will trigger payload eviction but overall NF processing will
stall because of lost connection to the NF server. Once connectivity is
restored, the space reclaimed by the payload evictor will resume normal
operation.
\newline
2) NF server failures: All in-flight packets through the
NF server will be lost in both deployments. NF chain processing will resume
after the NF server recovers. Similar to link failures, the payload evictor
will resume normal operation.
\newline
3) Switch failure: \PP{} increases the
failure-domain of the NF-server to include the switch. Therefore, when the switch
fails, all packets in the switch will be lost in both deployments. However,
the failure scenario differs for packets that have already reached the NF
server. In the baseline deployment,
if server connects to another ToR switch, it
can route packets through a different ToR.
But, this is not applicable to \PP{}, because the payload resides in
the original ToR switch memory.
However, the number of lost packets is negligible due
to small time-delta between the Split and Merge operations (that includes NF
processing). For example, in Fig.~\ref{fig:PeakGoodput}, we will lose at most 50 packets
for a link operating at full capacity (10 Gbps) with an average packet size of 882 bytes
and an average latency of 32 $\mu$s. With a 40 Gbps link, we will lose at most 200 packets. This packet loss is an overestimate, because
we measure latency from the traffic generator and not from the switch.
\newline
\textbf{Security.} We do not protect the switch and the stored payload against
adversarial attacks. We apply \PP{} in the context of the cloud deployment
where both the switch and the NF server are within the cloud provider's administrative domain.
Since the cloud provider owns the infrastructure, we trust all incoming
traffic to the switch.
\newline \noindent \textbf{Adaptive payload eviction policy.} Our prototype
tracks premature payload evictions with a counter.
In our evaluation, we use this counter to identify the safe operating boundary
of \PP{}.
This counter could be used to adaptively change the payload eviction policy
and protect against unexpected latency spikes in the NF server.
For example, \PP{} could start with an aggressive payload eviction policy and dynamically
switch to a conservative eviction policy
when payload evictions exceed a predefined threshold.
We leave this tuning to future work.

\section{Related Work}
\label{sec:related_work}

This is an extended version our conference article~\cite{pp_conext}.
Our work draws inspiration from Cut Payload proposed
by Cheng et al.~\cite{cutpayload}.
Cut Payload drops payload at overloaded switches and
generates SACK-style notifications to the sender about the drops.
\PP{} makes a similar observation about not transmitting unnecessary data.
Our approach is different because we do not drop the payload.
Instead, we store the payload at the switch and later merge it back with the header.
In addition, our work broadly spans two areas: in-network computing and NFs.

\noindent \textbf{In-network computing.}
Prior work has used in-network processors to accelerate applications by
offloading application functionality.
\PP{} differs by using the switch to implement a transparent optimization,
without modifying application code.
For example,
NetCache implements an in-network cache for key-value stores~\cite{netcache}. DistCache presents a distributed
cache topology to load balance across racks~\cite{DistCache}.
Silkroad offloads
L4 load balancers~\cite{Miao2017SilkRoad}, and NetPaxos~\cite{netpaxos} offloads
some Paxos functions to the switch.
NOPaxos (Network Ordered Paxos) uses in-network devices for network sequencing and
accelerates data replication~\cite{NOPaxos}. Researchers have also
used the switch dataplane to accelerate SQL queries (Jumpgate~\cite{jumpgate}, Cheetah~\cite{Cheetah}) and string searching (PPS~\cite{SoulePPS}).

\noindent \textbf{Specialized hardware for NFs.}
Prior work has used specialized hardware to improve NF performance.
Wu et al. propose NF acceleration using RMT switches
by deploying NF chains directly on the switch~\cite{dejavu}. Our work differs
in that \PP{} is a
\textit{transparent in-network optimization}, and the
NF chains continue to run on commodity hardware. This retains the flexibility of implementing and
upgrading NF chains, while ensuring ease of integration with existing NF frameworks.
Researchers have also used other specialized devices for NF acceleration.
For example, PacketShader uses GPUs for accelerating software routers~\cite{packetshader}.
G-NET presents virtualization and isolation approaches for GPU-accelerated
NF chains~\cite{GNET}.
GEN elastically scales NFs using GPUs~\cite{gen-gpu}.
NBA~\cite{NBA} and GPU-NFV~\cite{GPUNFV} use GPUs to accelerate NF performance.
Similarly, ClickNP~\cite{clickNP} uses FPGAs.
Moon et al. offload stateful flow processing to programmable NICs~\cite{Moon:2018:AFP:3265723.3265744}.
Metron integrates end-hosts and in-network processing by offloading packet
classification and stateless operations of NFs to the switch and programmable
NICs, resulting in better latency and throughput~\cite{metron}.

\noindent \textbf{End-host optimizations for NFs.}
There are several end-host NF frameworks that optimize NF performance, and
\PP{} can be integrated with these frameworks.
CoMB consolidates NFs to reduce provisioning and maintenance
costs~\cite{comb}. Sprayer optimizes utilization in multi-core CPUs by
using fine-grained packet-to-core allocation~\cite{sprayer}.
Katsikas et al. optimize NF execution by profiling packet processing~\cite{katsikas2017profiling}.
TNP optimizes allocation of NFs to multiple cores~\cite{tnp}.
mOS provides an API to help develop stateful flow processing
programs~\cite{mos}.
SafeBricks protects NFs in untrusted clouds, but at a performance cost~\cite{safebricks}.
NetBricks uses Rust to eliminate isolation overhead of Docker containers and
VMs~\cite{NetBricks}.
NFP~\cite{nfp} and Parabox~\cite{ParaBox} explore parallel
paths in execution of NF chains to reduce packet processing latency.
We can seamlessly integrate \PP{} with existing frameworks.
For example, in the combined setup composed of \PP{} and NFP, \PP{}
would improve goodput and NFP would improve latency.

\section{Conclusion}
\label{sec:conclusion}

We described PayloadPark, an in-network optimization that decouples
packets at the header-payload boundary.
PayloadPark improves shallow NF goodput by 2-36\% without any latency penalty,
reduces NF server's PCIe load by 2-58\%,
and uses less than 40\% of Tofino chip resources.
Also, \PP{} preserves the semantics of non-\PP{} deployments,
and can be easily integrated with existing NF frameworks.

\section{Acknowledgements}

We thank the anonymous reviewers and our CoNEXT'20 conference
shepherd for their insightful
and constructive feedback. We would also like to thank Bruce Shepherd,
Joel Nider, and William Anthony Mason for reviewing early drafts of the paper
and giving us valuable feedback.

\bibliographystyle{plain}
\bibliography{paper}

\begin{thebibliography}{10}

\bibitem{pp_code}
{PayloadPark on Github}, 2020.
\newblock https://github.com/PayloadPark/payloadpark.

\bibitem{dctraffic}
Theophilus Benson, Aditya Akella, and David~A. Maltz.
\newblock {Network Traffic Characteristics of Data Centers in the Wild}.
\newblock In {\em Proceedings of the 10th ACM SIGCOMM Conference on Internet
  Measurement}, IMC '10, pages 267--280. ACM, 2010.

\bibitem{Bosshart2014P4}
Pat Bosshart, Dan Daly, Glen Gibb, Martin Izzard, Nick McKeown, Jennifer
  Rexford, Cole Schlesinger, Dan Talayco, Amin Vahdat, George Varghese, and
  David Walker.
\newblock {P4: Programming Protocol-independent Packet Processors}.
\newblock {\em SIGCOMM Comput. Commun. Rev.}, 44(3):87--95, 2014.

\bibitem{rmt}
Pat Bosshart, Glen Gibb, Hun-Seok Kim, George Varghese, Nick McKeown, Martin
  Izzard, Fernando Mujica, and Mark Horowitz.
\newblock {Forwarding Metamorphosis: Fast Programmable Match-action Processing
  in Hardware for SDN}.
\newblock In {\em Proceedings of the 2013 Conference of the ACM Special
  Interest Group on Data Communication}, SIGCOMM '13, pages 99--110. ACM, 2013.

\bibitem{openbox}
Anat Bremler-Barr, Yotam Harchol, and David Hay.
\newblock {OpenBox: A Software-Defined Framework for Developing, Deploying, and
  Managing Network Functions}.
\newblock In {\em Proceedings of the 2016 ACM SIGCOMM Conference}, SIGCOMM '16,
  pages 511--524. ACM, 2016.

\bibitem{cutpayload}
Peng Cheng, Fengyuan Ren, Ran Shu, and Chuang Lin.
\newblock {Catch the Whole Lot in an Action: Rapid Precise Packet Loss
  Notification in Data Centers}.
\newblock In {\em Proceedings of the 11th USENIX Conference on Networked
  Systems Design and Implementation}, NSDI '14, pages 17--28. USENIX
  Association, 2014.

\bibitem{p416}
The P4~Language Consortium.
\newblock {P4 16 Language Specification}, 2017.
\newblock https://p4.org/p4-spec/docs/P4-16-v1.0.0-spec.pdf.

\bibitem{netpaxos}
Huynh~Tu Dang, Daniele Sciascia, Marco Canini, Fernando Pedone, and Robert
  Soul{\'e}.
\newblock {NetPaxos: Consensus at Network Speed}.
\newblock In {\em Proceedings of the Symposium on SDN Research}, SOSR '15,
  pages 1--7. ACM, 2015.

\bibitem{rdtsc}
DPDK.
\newblock {RDTSC API}, 2019.
\newblock \url{https://doc.dpdk.org/api-18.05/rte__cycles_8h.html}.

\bibitem{Maglev}
Daniel~E. Eisenbud, Cheng Yi, Carlo Contavalli, Cody Smith, Roman Kononov, Eric
  Mann-Hielscher, Ardas Cilingiroglu, Bin Cheyney, Wentao Shang, and
  Jinnah~Dylan Hosein.
\newblock {Maglev: A Fast and Reliable Software Network Load Balancer}.
\newblock In {\em Proceedings of the 13th USENIX Conference on Networked
  Systems Design and Implementation}, NSDI '16, pages 523--535. USENIX
  Association, 2016.

\bibitem{etsi-whitepaper}
ETSI.
\newblock {NFV Whitepaper}, 2013.
\newblock \url{https://portal.etsi.org/nfv/nfv_white_paper2.pdf}.

\bibitem{Fernandes2009LWDPI}
St\^{e}nio Fernandes, Rafael Antonello, Thiago Lacerda, Alysson Santos, Djamel
  Sadok, and Tord Westholm.
\newblock {Slimming Down Deep Packet Inspection Systems}.
\newblock In {\em Proceedings of the 28th IEEE International Conference on
  Computer Communications Workshops}, INFOCOM'09, pages 61--66. IEEE Press,
  2009.

\bibitem{pcm}
The~Linux Foundation.
\newblock {Process Counter Monitor}, 2011.
\newblock https://github.com/opcm/pcm.

\bibitem{pp_conext}
Swati Goswami, Nodir Kodirov, Craig Mustard, Ivan Beschastnikh, and Margo
  Seltzer.
\newblock {Parking Packet Payload with P4}.
\newblock 2020.

\bibitem{Han2015NFVChallenges}
B.~{Han}, V.~{Gopalakrishnan}, L.~{Ji}, and S.~{Lee}.
\newblock {Network function virtualization: Challenges and opportunities for
  innovations}.
\newblock {\em IEEE Communications Magazine}, 53(2):90--97, 2015.

\bibitem{packetshader}
Sangjin Han, Keon Jang, KyoungSoo Park, and Sue Moon.
\newblock {PacketShader: A GPU-accelerated Software Router}.
\newblock In {\em Proceedings of the 2010 Conference of the ACM Special
  Interest Group on Data Communication}, SIGCOMM '10, pages 195--206. ACM,
  2010.

\bibitem{FW_RFC}
IETF.
\newblock { Benchmarking Methodology for Firewall Performance}, 2003.
\newblock https://tools.ietf.org/html/rfc3511.

\bibitem{tofino_switch}
Intel.
\newblock {Tofino ASIC}, 2019.
\newblock
  https://www.intel.com/content/www/us/en/products/network-io/programmable-ethernet-switch/tofino-series/tofino.html.

\bibitem{mos}
Muhammad~Asim Jamshed, YoungGyoun Moon, Donghwi Kim, Dongsu Han, and KyoungSoo
  Park.
\newblock {mOS: A Reusable Networking Stack for Flow Monitoring Middleboxes}.
\newblock In {\em Proceedings of the 14th {USENIX} Symposium on Networked
  Systems Design and Implementation}, NSDI '17, pages 113--129. {USENIX}
  Association, 2017.

\bibitem{SoulePPS}
Theo Jepsen, Daniel Alvarez, Nate Foster, Changhoon Kim, Jeongkeun Lee, Masoud
  Moshref, and Robert Soul{\'e}.
\newblock {Fast String Searching on PISA}.
\newblock In {\em Proceedings of the Symposium on SDN Research}, SOSR '19,
  pages 21--28. ACM, 2019.

\bibitem{SpeedyBox}
Yimin Jiang, Yong Cui, Wenfei Wu, Zhe Xu, Jiahan Gu, K.~K. Ramakrishnan,
  Yongchao He, and Xuehai Qian.
\newblock {SpeedyBox: Low-Latency NFV Service Chains with Cross-NF Runtime
  Consolidation}.
\newblock In {\em International Conference on Distributed Computing Systems},
  ICDCS '19, pages 68--79. IEEE, 2019.

\bibitem{netcache}
Xin Jin, Xiaozhou Li, Haoyu Zhang, Robert Soul{\'e}, Jeongkeun Lee, Nate
  Foster, Changhoon Kim, and Ion Stoica.
\newblock {NetCache: Balancing Key-Value Stores with Fast In-Network Caching}.
\newblock In {\em Proceedings of the 26th Symposium on Operating Systems
  Principles}, SOSP '17, pages 121--136. ACM, 2017.

\bibitem{BurstRadar}
Raj Joshi, Ting Qu, Mun~Choon Chan, Ben Leong, and Boon~Thau Loo.
\newblock {BurstRadar: Practical Real-time Microburst Monitoring for Datacenter
  Networks}.
\newblock In {\em Proceedings of the 9th Asia-Pacific Workshop on Systems},
  APSys '18, pages 8:1--8:8. ACM, 2018.

\bibitem{metron}
Georgios~P. Katsikas, Tom Barbette, Dejan Kosti{\'c}, Rebecca Steinert, and
  Gerald Q.~Maguire Jr.
\newblock {Metron: {NFV} Service Chains at the True Speed of the Underlying
  Hardware}.
\newblock In {\em Proceedings of the 15th USENIX Conference on Networked
  Systems Design and Implementation}, NSDI '18, pages 171--186. {USENIX}
  Association, 2018.

\bibitem{katsikas2017profiling}
Georgios~P Katsikas, Gerald~Q Maguire~Jr, and Dejan Kosti{\'c}.
\newblock {Profiling and accelerating commodity NFV service chains with SCC}.
\newblock {\em Journal of Systems and Software}, 127:12--27, 2017.

\bibitem{NBA}
Joongi Kim, Keon Jang, Keunhong Lee, Sangwook Ma, Junhyun Shim, and Sue Moon.
\newblock {NBA (Network Balancing Act): A High-performance Packet Processing
  Framework for Heterogeneous Processors}.
\newblock In {\em Proceedings of the Tenth European Conference on Computer
  Systems}, EuroSys '15, pages 22:1--22:14. ACM, 2015.

\bibitem{clickNP}
Bojie Li, Kun Tan, Layong~(Larry) Luo, Yanqing Peng, Renqian Luo, Ningyi Xu,
  Yongqiang Xiong, Peng Cheng, and Enhong Chen.
\newblock {ClickNP: Highly Flexible and High Performance Network Processing
  with Reconfigurable Hardware}.
\newblock In {\em Proceedings of the 2016 Conference of the ACM Special
  Interest Group on Data Communication}, SIGCOMM '16, pages 1--14. ACM, 2016.

\bibitem{NOPaxos}
Jialin Li, Ellis Michael, Naveen~Kr. Sharma, Adriana Szekeres, and Dan R.~K.
  Ports.
\newblock {Just Say No to Paxos Overhead: Replacing Consensus with Network
  Ordering}.
\newblock In {\em Proceedings of the 12th USENIX Conference on Operating
  Systems Design and Implementation}, OSDI'16, pages 467--483. USENIX
  Association, 2016.

\bibitem{tnp}
M.~{Liu}, G.~{Feng}, J.~{Zhou}, and S.~{Qin}.
\newblock {Joint Two-Tier Network Function Parallelization on Multicore
  Platform}.
\newblock In {\em 2018 IEEE Global Communications Conference (GLOBECOM)}, pages
  1--7, 2018.

\bibitem{E3}
Ming Liu, Simon Peter, Arvind Krishnamurthy, and Phitchaya~Mangpo
  Phothilimthana.
\newblock {E3: Energy-Efficient Microservices on SmartNIC-Accelerated Servers}.
\newblock In {\em Proceedings of the 2019 USENIX Conference on Usenix Annual
  Technical Conference}, USENIX ATC '19, page 363–378. USENIX Association,
  2019.

\bibitem{DistCache}
Zaoxing Liu, Zhihao Bai, Zhenming Liu, Xiaozhou Li, Changhoon Kim, Vladimir
  Braverman, Xin Jin, and Ion Stoica.
\newblock {DistCache: Provable Load Balancing for Large-Scale Storage Systems
  with Distributed Caching}.
\newblock In {\em 17th {USENIX} Conference on File and Storage Technologies
  ({FAST} 19)}, pages 143--157. {USENIX} Association, 2019.

\bibitem{rfc7665}
Kari Marttila.
\newblock { Service Function Chaining (SFC) Architecture}, 2015.
\newblock https://tools.ietf.org/html/rfc7665.

\bibitem{McCauleyLoadBalancing}
James McCauley, Aurojit Panda, Arvind Krishnamurthy, and Scott Shenker.
\newblock {Thoughts on Load Distribution and the Role of Programmable
  Switches}.
\newblock {\em SIGCOMM Comput. Commun. Rev.}, 49(1):18–23, 2019.

\bibitem{Miao2017SilkRoad}
Rui Miao, Hongyi Zeng, Changhoon Kim, Jeongkeun Lee, and Minlan Yu.
\newblock {SilkRoad: Making Stateful Layer-4 Load Balancing Fast and Cheap
  Using Switching ASICs}.
\newblock In {\em Proceedings of the 2017 Conference of the ACM Special
  Interest Group on Data Communication}, SIGCOMM '17, pages 15--28. ACM, 2017.

\bibitem{Moon:2018:AFP:3265723.3265744}
Young~Gyoun Moon, Ilwoo Park, Seungeon Lee, and Kyoung~Soo Park.
\newblock {Accelerating Flow Processing Middleboxes with Programmable NICs}.
\newblock In {\em Proceedings of the 9th Asia-Pacific Workshop on Systems},
  APSys '18, pages 14:1--14:3. ACM, 2018.

\bibitem{jumpgate}
Craig Mustard, Fabian Ruffy, Anny Gakhokidze, Ivan Beschastnikh, and Alexandra
  Fedorova.
\newblock {Jumpgate: In-network Processing As a Service for Data Analytics}.
\newblock In {\em Proceedings of the 11th USENIX Conference on Hot Topics in
  Cloud Computing}, HotCloud'19, pages 6--6. USENIX Association, 2019.

\bibitem{CostOfHTTPS}
David Naylor, Alessandro Finamore, Ilias Leontiadis, Yan Grunenberger, Marco
  Mellia, Maurizio Munaf\`{o}, Konstantina Papagiannaki, and Peter Steenkiste.
\newblock {The Cost of the "S" in HTTPS}.
\newblock In {\em Proceedings of the 10th ACM International on Conference on
  Emerging Networking Experiments and Technologies}, CoNEXT '14, pages
  133--140. ACM, 2014.

\bibitem{pcie_bench}
Rolf Neugebauer, Gianni Antichi, Jos\'{e}~Fernando Zazo, Yury Audzevich, Sergio
  L\'{o}pez-Buedo, and Andrew~W. Moore.
\newblock {Understanding PCIe Performance for End Host Networking}.
\newblock In {\em Proceedings of the 2018 Conference of the ACM Special
  Interest Group on Data Communication}, SIGCOMM '18, page 327–341. ACM,
  2018.

\bibitem{NetBricks}
Aurojit Panda, Sangjin Han, Keon Jang, Melvin Walls, Sylvia Ratnasamy, and
  Scott Shenker.
\newblock {NetBricks: Taking the V out of NFV}.
\newblock In {\em Proceedings of the 12th USENIX Conference on Operating
  Systems Design and Implementation}, OSDI'16, pages 203--216. USENIX
  Association, 2016.

\bibitem{ananta}
Parveen Patel, Deepak Bansal, Lihua Yuan, Ashwin Murthy, Albert Greenberg,
  David~A. Maltz, Randy Kern, Hemant Kumar, Marios Zikos, Hongyu Wu, Changhoon
  Kim, and Naveen Karri.
\newblock {Ananta: Cloud Scale Load Balancing}.
\newblock {\em SIGCOMM Comput. Commun. Rev.}, 43(4):207–218, 2013.

\bibitem{safebricks}
Rishabh Poddar, Chang Lan, Raluca~Ada Popa, and Sylvia Ratnasamy.
\newblock {SafeBricks: Shielding Network Functions in the Cloud}.
\newblock In {\em Proceedings of the 15th {USENIX} Symposium on Networked
  Systems Design and Implementation}, NSDI '18, pages 201--216. {USENIX}
  Association, 2018.

\bibitem{sprayer}
Hugo Sadok, Miguel Elias~M. Campista, and Lu\'{\i}s Henrique M.~K. Costa.
\newblock {A Case for Spraying Packets in Software Middleboxes}.
\newblock In {\em Proceedings of the 17th ACM Workshop on Hot Topics in
  Networks}, HotNets '18, pages 127--133. ACM, 2018.

\bibitem{comb}
Vyas Sekar, Norbert Egi, Sylvia Ratnasamy, Michael~K. Reiter, and Guangyu Shi.
\newblock {Design and Implementation of a Consolidated Middlebox Architecture}.
\newblock In {\em Proceedings of the 9th USENIX Conference on Networked Systems
  Design and Implementation}, NSDI '12, pages 24--24. USENIX Association, 2012.

\bibitem{APLOMB}
Justine Sherry, Shaddi Hasan, Colin Scott, Arvind Krishnamurthy, Sylvia
  Ratnasamy, and Vyas Sekar.
\newblock {Making Middleboxes Someone else's Problem: Network Processing As a
  Cloud Service}.
\newblock In {\em Proceedings of the 2012 Conference of the ACM Special
  Interest Group on Data Communication}, SIGCOMM '12, pages 13--24. ACM, 2012.

\bibitem{nfp}
Chen Sun, Jun Bi, Zhilong Zheng, Heng Yu, and Hongxin Hu.
\newblock {NFP: Enabling Network Function Parallelism in NFV}.
\newblock In {\em Proceedings of the 2017 Conference of the ACM Special
  Interest Group on Data Communication}, SIGCOMM '17, pages 43--56. ACM, 2017.

\bibitem{Cheetah}
Muhammad Tirmazi, Ran Ben~Basat, Jiaqi Gao, and Minlan Yu.
\newblock {Cheetah: Accelerating Database Queries with Switch Pruning}.
\newblock In {\em Proceedings of the ACM SIGCOMM 2019 Conference Posters and
  Demos}, SIGCOMM Posters and Demos '19, pages 72--74. ACM, 2019.

\bibitem{ResQ}
Amin Tootoonchian, Aurojit Panda, Chang Lan, Melvin Walls, Katerina Argyraki,
  Sylvia Ratnasamy, and Scott Shenker.
\newblock {ResQ: Enabling SLOs in Network Function Virtualization}.
\newblock In {\em Proceedings of the 15th {USENIX} Symposium on Networked
  Systems Design and Implementation}, NSDI '18, pages 283--297. {USENIX}
  Association, 2018.

\bibitem{caida-dataset}
The~CAIDA UCSD.
\newblock {nyc (dirA) - 2019-01-17}, 2020.
\newblock
  \url{https://www.caida.org/data/passive/trace_stats/nyc-A/2019/equinix-nyc.dirA.20190117-130000.UTC.df.txt}.

\bibitem{dejavu}
Dingming Wu, Ang Chen, T.~S.~Eugene Ng, Guohui Wang, and Haiyong Wang.
\newblock {Accelerated Service Chaining on a Single Switch ASIC}.
\newblock In {\em Proceedings of the 18th ACM Workshop on Hot Topics in
  Networks}, HotNets '19, page 141–149. ACM, 2019.

\bibitem{GPUNFV}
Xiaodong Yi, Jingpu Duan, and Chuan Wu.
\newblock {GPUNFV: A GPU-Accelerated NFV System}.
\newblock In {\em Proceedings of the First Asia-Pacific Workshop on
  Networking}, APNet'17, pages 85--91. ACM, 2017.

\bibitem{GNET}
Kai Zhang, Bingsheng He, Jiayu Hu, Zeke Wang, Bei Hua, Jiayi Meng, and Lishan
  Yang.
\newblock {G-NET: Effective {GPU} Sharing in {NFV} Systems}.
\newblock In {\em Proceedings of the 15th {USENIX} Symposium on Networked
  Systems Design and Implementation}, NSDI '18, pages 187--200. {USENIX}
  Association, 2018.

\bibitem{opennetvm}
Wei Zhang, Guyue Liu, Wenhui Zhang, Neel Shah, Phillip Lopreiato, Gregoire
  Todeschi, K.K. Ramakrishnan, and Timothy Wood.
\newblock {OpenNetVM: A Platform for High Performance Network Service Chains}.
\newblock In {\em Proceedings of the 2016 Workshop on Hot Topics in Middleboxes
  and Network Function Virtualization}, HotMiddlebox '16, pages 26--31. ACM,
  2016.

\bibitem{ParaBox}
Yang Zhang, Bilal Anwer, Vijay Gopalakrishnan, Bo~Han, Joshua Reich, Aman
  Shaikh, and Zhi-Li Zhang.
\newblock {ParaBox: Exploiting Parallelism for Virtual Network Functions in
  Service Chaining}.
\newblock In {\em Proceedings of the Symposium on SDN Research}, SOSR '17,
  pages 143--149. ACM, 2017.

\bibitem{gen-gpu}
Zhilong Zheng, Jun Bi, Chen Sun, Heng Yu, Hongxin Hu, Zili Meng, Shuhe Wang,
  Kai Gao, and Jianping Wu.
\newblock {GEN: A GPU-Accelerated Elastic Framework for NFV}.
\newblock In {\em Proceedings of the 2Nd Asia-Pacific Workshop on Networking},
  APNet '18, pages 57--64. ACM, 2018.

\end{thebibliography}

\end{document}